\documentclass[journal]{IEEEtran}
\usepackage{amsmath,amsfonts}
\newtheorem{theorem}{Theorem}
\allowdisplaybreaks[1] 
\usepackage{algorithmic}
\usepackage{algorithm}
\usepackage{array}
\usepackage[export]{adjustbox} 
\usepackage{textcomp}
\usepackage{color} 
\usepackage{stfloats}
\usepackage{url}
\usepackage{xcolor} 
\usepackage{verbatim}
\usepackage{graphicx}
\usepackage{cite}
\usepackage{bm}  
\usepackage{subfigure} 
\newtheorem{lemma}{Lemma}
\begin{document}
	
	\title{Geometry-based Efficient Airy Beam Training for Quasi-LoS in Terahertz Near-field Communications}
	
	\author{Wenqi Zhao and Chong Han,~\IEEEmembership{Senior Member,~IEEE} 
		\thanks{
			A preliminary and shorter version of this paper appeared in the arXiv preprint available at https://arxiv.org/abs/2605.09895. 
			
			Wenqi Zhao is with the Terahertz Wireless Communications Interdisciplinary Research Center (TWC IRC), Shanghai Jiao Tong University, Shanghai 200240,
			China (e-mail: wenqi.zhao@sjtu.edu.cn).
			
			Chong Han is with the Terahertz Wireless Communications Interdisciplinary Research Center (TWC IRC) and also the Cooperative Medianet Innovation Center (CMIC), School of Information Science and Electronic Engineering, Shanghai Jiao Tong University, Shanghai 200240,
			China (e-mail: chong.han@sjtu.edu.cn).

	}}
	
	
	
	
	\maketitle
	\thispagestyle{empty}
	
	\begin{abstract}
		With the enlargement of antenna apertures in Terahertz (THz) communications, near-field propagation becomes increasingly dominant. Moving beyond conventional line-of-sight (LoS) and non-line-of-sight (NLoS) scenarios, quasi-line-of-sight (quasi-LoS) communications with partial blockage have emerged as an important propagation regime. Airy beams provide a promising solution for blockage avoidance due to their unique curved propagation trajectories. However, when the blockage location is unknown, existing Airy beam training methods mainly rely on parameter-domain search or learning-based approaches, leading to high training overhead and limited geometric interpretability. In this paper, we propose a geometry-based efficient Airy beam training framework for both uniform linear array (ULA) and uniform planar array (UPA) systems. First, the theoretical feasibility bounds of Airy beam generation under finite apertures are derived to eliminate physically infeasible codewords. Based on these bounds, a performance-oriented Slope-Depth Geometric Codebook (SDGC) is developed by combining bending-direction probing with geometric sampling in the slope-depth domain. Furthermore, a latency-oriented Bijective 1D Geometric Codebook (B1GC) is proposed by establishing a bijective mapping between the blockage-avoidance waypoint and the receiver target, thereby reducing the beam training process to a one-dimensional geometric search. The framework is further extended to UPA systems through a hybrid focusing--Airy mode, where four probe beams jointly determine the transverse dimension for blockage avoidance and the corresponding bending direction, followed by Airy beam training along the selected dimension and near-field focusing along the orthogonal dimension. Simulation results show that SDGC achieves an average spectral efficiency of 14.8 bit/s/Hz while reducing the training overhead by 40.9\% compared with the hierarchical focusing--Airy codebook (HFAC), whereas B1GC reduces the overhead by 87.1\% while maintaining comparable performance. The UPA results further verify the effectiveness of the proposed framework under two-dimensional blockage scenarios.
	\end{abstract}
	
	
	\section{Introduction}
	
	\IEEEPARstart{W}{ith} the rapid evolution of future wireless communication systems, the demands for ultra-high data rates and ultra-low latency continue to grow dramatically~\cite{Akyildiz-2022-Terahertz,Chen-2019-A-survey}. Driven by emerging applications such as AI-native services and wireless data centers, future networks are expected to require terabit-per-second (Tbps) wireless connectivity. To support such unprecedented capacity demands, the Terahertz (THz) band (0.1–10~THz) has emerged as a key enabling technology due to its abundant spectrum resources spanning tens of GHz or even wider bandwidths. To compensate for the free-space attenuation and molecular absorption loss at THz frequencies, ultra-massive multiple-input multiple-output (UM-MIMO) systems employing hundreds or even thousands of antennas are widely adopted to generate highly directional beams with large array gains~\cite{Han-2021-Hybrid}. 
	As the operating frequency and array aperture increase, the Rayleigh distance expands significantly, causing THz communications to predominantly operate in the near-field region rather than the conventional far-field region~\cite{An-2024-Near-Field}. Consequently, the electromagnetic wave model transitions from plane-wave propagation to spherical-wave propagation, thereby introducing new spatial degrees of freedom for advanced near-field wavefront shaping.

	Despite the enormous potential of THz communications, THz links remain vulnerable to blockage by surrounding obstacles. In practical communication environments, the enlarged array aperture and near-field propagation characteristics make the channel conditions no longer limited to ideal line-of-sight (LoS) or fully blocked non-line-of-sight (NLoS) scenarios. Instead, partially blocked LoS environments, referred to as quasi-LoS scenarios~\cite{Yuan-2023-Spatial}, become increasingly common, especially in complex indoor deployments such as wireless data centers where dense server racks, structural
	walls, and metallic mesh antenna arrays substantially raise the probability of partial LoS blockage, with up to 52\% partial obstruction~\cite{Han-2025-datacenter-mag}. Consequently, quasi-LoS propagation, lying between conventional LoS and NLoS conditions, emerges as an important communication scenario for future THz networks.
	
	However, existing beam patterns suffer from severe limitations under quasi-LoS conditions. Conventional far-field steering beams and near-field focusing beams, which fundamentally belong to Gaussian beams with Gaussian-distributed field profiles~\cite{Ayach-2014-Spatially,Alkhateeb-2015-Limited,Zhang-2022-Beam-focusing,Cui-2022-Channel-estimation,Cui-2023-Near-Field-MIMO,Wu-2023-Multiple}, perform well under LoS propagation but experience substantial performance degradation once the LoS path is partially blocked, due to their limited ability to bypass obstacles. In contrast, Airy beams, characterized by cubic phase modulation and Airy-function-based transverse field distributions, exhibit unique self-acceleration and non-diffraction propagation properties~\cite{Nikolaos-2019-Airy}. Specifically, the self-acceleration characteristic enables the beam trajectory to bend around obstacles, while the non-diffraction property allows the beam to maintain concentrated energy over extended propagation distances. In addition, Airy beams exhibit self-healing capability, allowing the wavefront to partially reconstruct after encountering blockage. These unique propagation characteristics make Airy beams a highly promising solution for robust quasi-LoS THz near-field communications.

	\subsection{Related Work}
	
	\subsubsection{Airy Beam}
	
	The Airy beam, originally introduced in optics~\cite{Zhan-2020-Propagations,Nikolaos-2019-Airy}, has recently attracted increasing attention in THz communications~\cite{Jornet-2023-Wireless} due to its curved propagation trajectory and blockage-bypassing capability.
	Early studies mainly focus on demonstrating the physical feasibility of curved THz propagation. Experimental results in~\cite{Guerboukha-2024-Curving} show that high-data-rate THz links can be established around obstacles using Airy beams, while~\cite{Chen-2025-physics} further demonstrates practical Airy beam generation using a metasurface platform.
	Subsequent studies focus on the design and physical understanding of Airy wavefronts for blockage avoidance. In~\cite{Chen-2024-Curving}, a physics-informed learning framework optimizes the transmit phase profile to generate desired Airy trajectories. The convex-optimization-based bending beamforming method in~\cite{Liu-2025-Near-Field-THz-Bending} further enhances blockage robustness by optimizing the minimum received power along a prescribed propagation trajectory. In~\cite{Lu-2026-Analytical}, an analytical framework characterizes practical Airy beam trajectories under finite apertures and quantifies their performance tradeoffs with conventional focusing beams. More recently, \cite{Wang-2026-Cubic} investigates the physical mechanism underlying Airy blockage recovery from a phase-compensation perspective and shows that the linear and quadratic phase components compensate most of the blockage-induced phase mismatch, while the cubic Airy term provides the first independent correction to the remaining non-quadratic mismatch.
	
	To further simplify Airy beam configuration, our previous work~\cite{Zhao-2026-Airy} establishes a closed-form mapping from the geometric locations of the blockage and receiver to the Airy beam parameters, enabling direct geometry-based wavefront design without iterative parameter optimization. In~\cite{Wang-2026-Physics-Guided}, a physics-guided neural Airy beamforming method is developed, where a compact trajectory region is characterized from the sensed blockage geometry and a lightweight neural predictor selects a near-optimal waypoint, which is then converted into the Airy beam parameters through an analytical generation operator. Meanwhile, Airy beamforming also extends to more general scenarios. Beyond blockage avoidance, wavefront hopping dynamically switches wavefronts to enhance physical-layer security and mitigate interference~\cite{Petrov-2024-Wavefront,Petrov-2024-Wavefront-Hopping,Singh-2024-Wavefront}.
	
	\subsubsection{Airy Beam Training}
	
	When the blockage location is unknown, Airy beam training is required to establish reliable communication links. Existing approaches mainly rely on learning-based beam selection or search-based beam training. In~\cite{Weng-2025-Learning}, an approximate Airy beam trajectory model is derived, and a learning-based beam training framework using discrete Fourier transform (DFT) codebooks identifies suitable beam configurations under unknown blockage conditions. For practical THz applications, our prior work in~\cite{Zhao-2025-WDC} introduces a hierarchical focusing--Airy beam search scheme (HFAC), where the optimal focusing distance-angle pair is first determined, followed by a refinement stage for the curving parameter. In~\cite{zhang-2026-breaking}, a near-field Airy beam codebook is constructed by jointly discretizing the focusing, steering, trajectory-curvature, and amplitude-shaping parameters, and a low-overhead hierarchical search scheme is developed to identify suitable Airy beams under unknown obstacle and user locations.
	Furthermore, the authors in \cite{zhang-2026-hybrid} extend Airy beam training to multi-user near-field communications by designing a predefined Airy beam codebook and a multi-user hierarchical search framework. The selected Airy beams are subsequently employed to configure the hybrid analog beamformer for blockage avoidance and user alignment, while digital precoding is used to mitigate inter-user interference. Although these studies provide different solutions for Airy beam training, several limitations remain. Learning-based approaches require offline training and depend on the available training data and blockage-related input information. Search-based approaches mainly perform exhaustive or hierarchical sampling over abstract beam parameters without explicitly exploiting the geometric characteristics and physical trajectory properties of Airy beam propagation. As a result, the generated codewords may not efficiently align with the actual blockage geometry, leading to unnecessary beam training overhead. Moreover, existing Airy beam training studies mainly focus on ULA systems, while Airy beam training for UPA systems receives limited attention.
	
	\subsection{Contributions of Our Work}
	
	To address these challenges, building upon our previous work~\cite{Zhao-2026-Airy}, where a closed-form Airy beamforming framework is developed to directly determine the Airy beam parameters from geometric coordinates, this paper proposes a geometry-based efficient Airy beam training framework for both ULA and UPA systems. First, the theoretical feasibility boundary of Airy beam generation under finite aperture is derived to eliminate physically infeasible codewords and reduce beam training overhead. Based on this boundary, we develop a performance-oriented Slope-Depth Geometric Codebook (SDGC) and a latency-oriented Bijective 1D Geometric Codebook (B1GC), providing two complementary solutions for efficient Airy beam training with different performance--latency tradeoffs. The proposed framework is further extended to UPA systems by adopting the hybrid focusing--Airy mode in~\cite{Zhao-2026-Airy}, where probing is first employed to identify the transverse dimension for blockage avoidance, followed by Airy beam training along the selected dimension and near-field focusing along the orthogonal dimension. The main contributions are summarized as follows.
	
	\begin{itemize}
		
		\item \textbf{We derive the theoretical feasibility boundary of Airy beam generation under finite transmit aperture.}
		Specifically, by leveraging the caustic interpretation of Airy beam propagation, we establish a closed-form tangent-intercept criterion to characterize the physically realizable region of Airy beam trajectories. Based on this analysis, an explicit geometric boundary in the spatial domain is derived, which enables the proposed framework to efficiently prune physically infeasible codewords before beam training. This theoretical result provides both physical interpretability and practical guidance for low-overhead Airy beam codebook design.
		
		\item \textbf{We propose the SDGC, a performance-oriented 2D geometry-based Airy beam training framework based on slope-depth sampling.}
		The proposed framework first determines the bending direction of the Airy beam through a lightweight probing mechanism and then uniformly samples candidate waypoints in the transformed slope-depth domain within the feasible region, which induces a geometry-adaptive non-uniform distribution in the Cartesian spatial domain. By leveraging the closed-form relationship between the waypoint geometry and Airy beam parameters, SDGC achieves high spectral efficiency with significantly reduced training overhead.
		
		\item \textbf{We further develop the B1GC, a latency-oriented 1D geometry-based Airy beam training framework for rapid link establishment.}
		By fixing the reference depth and establishing a bijective mapping between the blockage-avoidance waypoint and the receiver target, B1GC converts the original 2D search space into a low-complexity 1D beam training process. Furthermore, a spatial coverage guarantee is established to ensure continuous coverage of the quasi-LoS region without blind spots. As a result, B1GC achieves ultra-low training latency while maintaining robust communication performance.
		
		\item \textbf{We extend the proposed geometry-based Airy beam training framework to UPA systems.}
		By adopting the hybrid focusing--Airy mode introduced in our previous work~\cite{Zhao-2026-Airy}, four hybrid probe beams are used to identify the transverse dimension for blockage avoidance and the corresponding bending direction. The subsequent SDGC or B1GC training is performed only along the selected dimension, while near-field focusing is retained along the orthogonal dimension. Simulation results in both ULA and UPA systems demonstrate that the proposed framework provides high spectral efficiency with low training overhead and maintains robust communication under blockage.
		
	\end{itemize}
	
	The remainder of the paper is organized as follows. The system model and the closed-form Airy beam framework are presented in Sec.~\ref{sec:system_model}. In Sec.~\ref{sec:Airy_boundary}, the theoretical boundary of Airy beam generation under finite array apertures is investigated. Then, the SDGC and B1GC are proposed in Sec.~\ref{sec:non-uniform_codebook_design} and Sec.~\ref{sec:fast scanning}, respectively. The extension of the proposed framework to UPA systems is presented in Sec.~\ref{sec:UPA}. Numerical evaluations are provided in Sec.~\ref{sec:simulation_results}, and the paper is concluded in Sec.~\ref{sec:conclusion}.

	\section{System Model}
	\label{sec:system_model}
	
	In this section, we introduce the communication system model for both ULA and UPA systems and review the closed-form Airy beam design framework developed in our previous work~\cite{Zhao-2026-Airy}. These results serve as the foundation for the geometry-based Airy beam training schemes developed in the subsequent sections.
	
	\subsection{Communication System}
	
	We consider a point-to-point THz near-field communication system in a three-dimensional Cartesian coordinate system, where the signal propagates primarily along the positive $z$-axis. The transmitter (Tx) is located on the plane $z=0$, while the receiver (Rx) is placed parallel to the Tx at a propagation distance $z=D$. The Tx and Rx arrays are centered at $(0,0,0)$ and $(0,0,D)$, respectively, with a uniform inter-element spacing $d$.
	
	For the general UPA system, the Tx consists of $N_{t,x}\times N_{t,y}$ antenna elements on the $x$-$y$ plane. The physical position of the $(n_x,n_y)$-th Tx element is
	\begin{equation}
		\mathbf{p}_{n_x,n_y}^{t}
		=
		\left[
		x_{n_x}^{t},
		y_{n_y}^{t},
		0
		\right]^{\mathrm T},
	\end{equation}
	where
	\begin{equation}
		x_{n_x}^{t}
		=
		\left(
		n_x-\frac{N_{t,x}-1}{2}
		\right)d,
		\qquad
		y_{n_y}^{t}
		=
		\left(
		n_y-\frac{N_{t,y}-1}{2}
		\right)d.
	\end{equation}
	Similarly, the position of the $(m_x,m_y)$-th Rx element is
	\begin{equation}
		\mathbf{p}_{m_x,m_y}^{r}
		=
		\left[
		x_{m_x}^{r},
		y_{m_y}^{r},
		D
		\right]^{\mathrm T},
	\end{equation}
	where
	\begin{equation}
		x_{m_x}^{r}
		=
		\left(
		m_x-\frac{N_{r,x}-1}{2}
		\right)d,
		\qquad
		y_{m_y}^{r}
		=
		\left(
		m_y-\frac{N_{r,y}-1}{2}
		\right)d.
	\end{equation}
	Here, $n_x=0,\ldots,N_{t,x}-1$, $n_y=0,\ldots,N_{t,y}-1$, $m_x=0,\ldots,N_{r,x}-1$, and $m_y=0,\ldots,N_{r,y}-1$.
	
	For the ULA system, we set $N_{t,y}=N_{r,y}=1$, such that the system reduces to the two-dimensional $x$-$z$ plane. We then denote $N_t=N_{t,x}$ and $N_r=N_{r,x}$, with array apertures $D_t=(N_t-1)d$ and $D_r=(N_r-1)d$. 
	\subsection{Closed-form Airy Beam Framework}
	
	According to the principle of Fourier optics~\cite{Nikolaos-2019-Airy}, an Airy beam can be generated by imposing a cubic phase profile on the transmit aperture. Taking the $x$-dimension as an example, the 1D phase profile is given by~\cite{Chen-2024-Curving,Chen-2025-physics}
	\begin{equation}
		\phi_x(x_0)
		=
		\frac{1}{3}(2\pi B)^3 x_0^3
		-
		\frac{\pi}{\lambda F}x_0^2
		-
		\frac{2\pi}{\lambda}\sin{\theta}x_0,
		\label{eq:airy_phase}
	\end{equation}
	where $x_0$ denotes the transverse coordinate of the transmit array, $\lambda$ is the wavelength, $B$ controls the bending behavior, $F$ denotes the focal distance, and $\theta$ represents the steering angle. The same formulation can be directly applied to the $y$-dimension for UPA systems.
	
	The trajectory of the Airy beam main lobe is derived in our previous work~\cite{Zhao-2026-Airy} as
	\begin{equation}
		\label{eq:traj}
		x(z)
		=
		-\xi\lambda z B
		-\sin{\theta}z
		-\frac{S_R^2-S_I^2}{16\lambda\pi^2B^3}z,
	\end{equation}
	where $\xi$ is the independent variable of the Airy function,
	$S_R=\frac{1}{z}-\frac{1}{F}$,
	$S_I=\frac{\lambda}{\pi\omega_0^2}$,
	and the Gaussian beam waist is $\omega_0=D_t/2$.
	The first main-lobe peak of the Airy function occurs at $\xi_1=-1.019$, and substituting $\xi=\xi_1$ into \eqref{eq:traj} yields the trajectory of the Airy beam main lobe.
	
	\begin{figure}[t]
		\centering
		\includegraphics[width=0.8\linewidth]{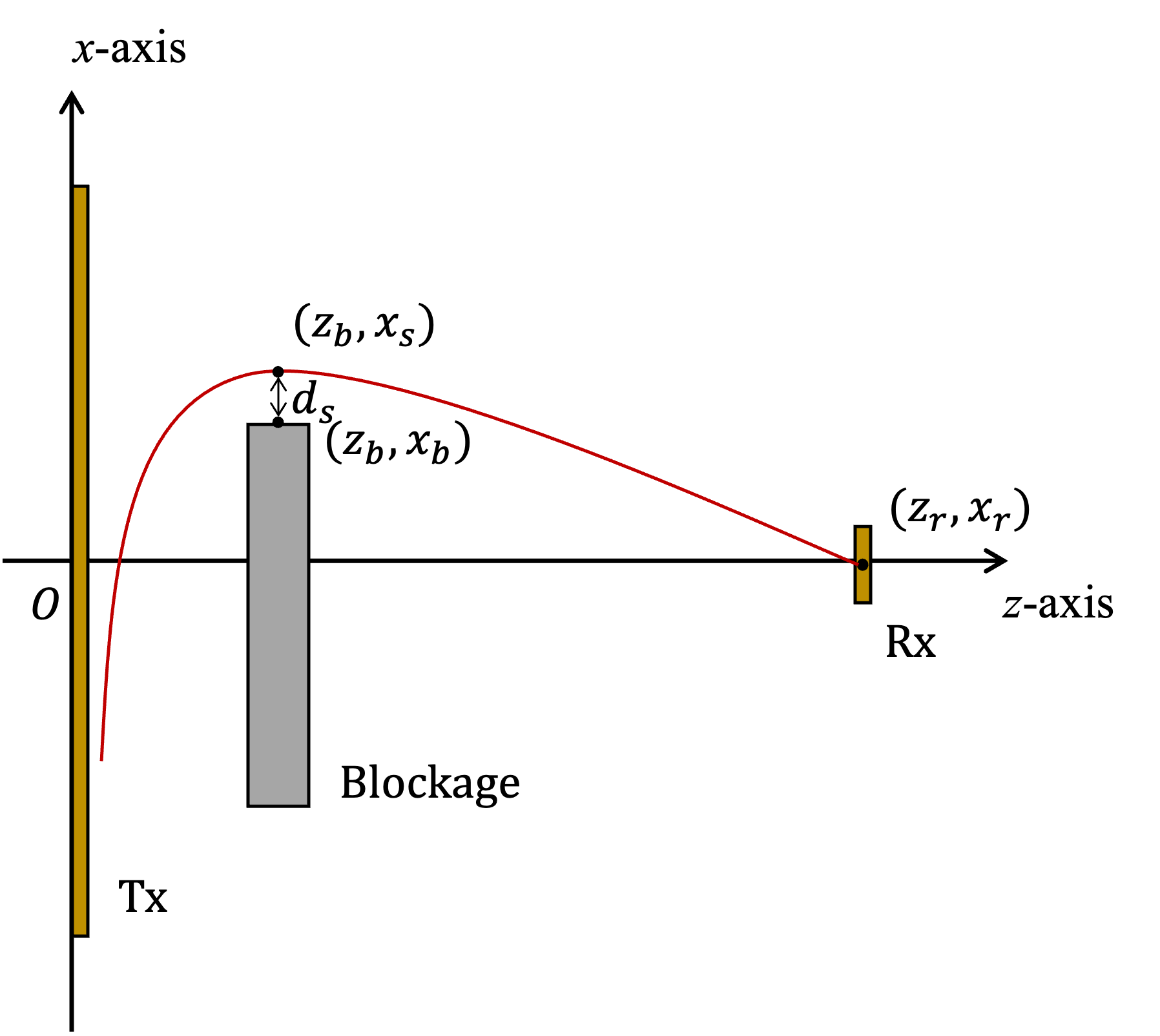}
		\caption{Closed-form Airy beam design in a ULA system.}
		\label{fig:system_airy}
	\end{figure}
	
	\begin{figure*}[!t]
		\normalsize
		
		\begin{equation} B_{{opt}} = \sqrt[3]{ -\frac{3 \left( \frac{x_r}{z_r} - \frac{x_s}{z_b} \right)}{16 \lambda \pi^2 \omega_0^2} + \sigma \sqrt{ \left[ \frac{3 \left( \frac{x_r}{z_r} - \frac{x_s}{z_b} \right)}{16 \lambda \pi^2 \omega_0^2} \right]^2 + \frac{2}{(2\pi)^6 \omega_0^6} + \frac{3 \left( \frac{1}{z_r} - \frac{1}{z_b} \right)^2}{128 \lambda^2 \pi^4 \omega_0^2} } } \label{eq:closed_form_B}  \end{equation}
		
		\begin{equation}
			F_{opt} = \frac{1}{ \frac{1}{2}\left(\frac{1}{z_r}+\frac{1}{z_b}\right) + \frac{ 8\lambda\pi^2 (\frac{x_r}{z_r} - \frac{x_s}{z_b}) }{ (\frac{1}{z_r} - \frac{1}{z_b}) } B^3 } \label{eq:closed_form_F}
		\end{equation}
		
		\begin{equation}
			\theta_{opt} = \arcsin\left( -\xi_{peak}\lambda B - \frac{x_s}{z_b} - \frac{ \left( \frac{1}{z_b} - \frac{1}{F} \right)^2 - \left(\frac{\lambda}{\pi\omega_0^2}\right)^2 }{ 16\lambda\pi^2 B^3 } \right)  \label{eq:closed_form_theta}
		\end{equation}
		
		\hrulefill
		\vspace*{4pt}
	\end{figure*}
	
	As shown in Fig.~\ref{fig:system_airy}, when the blockage waypoint $(z_b,x_s)$ and the receiver target $(z_r,x_r)$ are specified, the three Airy beam parameters $\{B,F,\theta\}$ can be directly obtained from the closed-form solutions developed in~\cite{Zhao-2026-Airy}, as given in \eqref{eq:closed_form_B}--\eqref{eq:closed_form_theta}. Here, $\sigma$ determines the bending branch, $(z_b,x_s)$ denotes the waypoint for blockage avoidance, and $(z_r,x_r)$ denotes the target point at the receiver.
	
	Let $\mathcal{A}(\cdot)$ denote the closed-form mapping from the prescribed geometric coordinates to the Airy beam parameters. The resulting Airy beamforming vector is expressed as
	\begin{equation}
		\mathbf{f}^{\mathrm{A}}
		=
		\mathcal{F}
		\left(
		\mathcal{A}(z_b,x_s;z_r,x_r)
		\right),
		\label{eq:closed_form_mapping}
	\end{equation}
	where $\mathcal{F}(\cdot)$ maps the derived parameters $\{B,F,\theta\}$ to the phase profile across the transmit array.
	
	For UPA systems, the phase profile can be independently constructed along the two transverse dimensions based on the separable formulation in~\cite{Zhao-2026-Airy}. In this work, we adopt the hybrid focusing--Airy mode, where Airy beamforming is applied along the transverse dimension used for blockage avoidance, while conventional near-field focusing is retained along the orthogonal dimension. Accordingly, the UPA beamforming vector is expressed as
	\begin{equation}
		\mathbf{f}^{\mathrm{UPA}}
		=
		\begin{cases}
			\mathbf{f}_{x}^{\mathrm{A}}\otimes\mathbf{f}_{y}^{\mathrm{F}},
			& \text{$x$-dimension blockage avoidance},\\[1mm]
			\mathbf{f}_{x}^{\mathrm{F}}\otimes\mathbf{f}_{y}^{\mathrm{A}},
			& \text{$y$-dimension blockage avoidance},
		\end{cases}
		\label{eq:upa_hybrid_mode}
	\end{equation}
	where $\mathbf{f}_{x}^{\mathrm{A}}$ and $\mathbf{f}_{y}^{\mathrm{A}}$ denote the Airy beamforming vectors along the $x$- and $y$-dimensions, respectively, and $\mathbf{f}_{x}^{\mathrm{F}}$ and $\mathbf{f}_{y}^{\mathrm{F}}$ denote the corresponding focusing vectors.
	\section{Theoretical Bounds of Airy Beam Generation under Finite Array Apertures}
	\label{sec:Airy_boundary}
	
	In this section, we characterize the fundamental feasibility bound imposed by the finite transmit aperture on Airy beam generation. This bound plays a central role in the subsequent codebook design, since it explicitly identifies the physically realizable region of Airy beam trajectories and enables the beam training framework to discard invalid beam candidates in advance, thereby significantly reducing the pilot overhead.
	
	For a given waypoint above the blockage and a target receiver position, the closed-form design in Sec.~\ref{sec:system_model} determines a set of beam parameters $\{B,F,\theta\}$ such that the main lobe of the Airy beam follows the prescribed trajectory. However, not every nominal trajectory can be physically generated by a finite-size transmit array. From the perspective of caustic theory, the intense curved main lobe of the Airy beam can be interpreted as the envelope of a family of constituent rays. Consequently, the tangent line to the trajectory at each propagation distance corresponds to one ray required to sustain the caustic structure. If the backward extension of this tangent line intersects the transmit plane outside the physical aperture, the corresponding ray cannot be launched by the array. In this case, aperture truncation occurs, which destroys the integrity of the caustic envelope and severely degrades the Airy beam trajectory. Therefore, the tangent intercept on the transmit plane provides a necessary physical realizability condition for Airy beam generation.
	
	To make this condition explicit, we first rewrite the trajectory of the  main lobe into a compact form. We define $K_1 = -\xi_{1}\lambda B - \sin\theta$ and
	$K_2 = \frac{1}{16\lambda\pi^2 B^3}$, where $\xi_{1}$ denotes the location of the first maximum point of the Airy function. Based on \eqref{eq:traj}, the trajectory can be expressed as
	\begin{equation}
		x(z)=\left(K_1+K_2S_I^2\right)z
		-
		K_2z\left(\frac{1}{z}-\frac{1}{F}\right)^2.
		\label{eq:xz_compact_revised}
	\end{equation}
	Taking the derivative of \eqref{eq:xz_compact_revised} with respect to $z$ yields
	\begin{equation}
		x'(z)
		=
		\left(K_1+K_2S_I^2\right)
		-
		K_2\left(-\frac{1}{z^2}+\frac{1}{F^2}\right).
		\label{eq:xprime_revised}
	\end{equation}
	Let $X_{\mathrm{int}}(z)$ denote the intercept between the tangent line of the trajectory and the transmit plane $z=0$. It is given by
	\begin{equation}
		X_{\mathrm{int}}(z)
		=
		x(z)-zx'(z).
		\label{eq:Xint_definition}
	\end{equation}
	Substituting \eqref{eq:xz_compact_revised} and \eqref{eq:xprime_revised} into \eqref{eq:Xint_definition} gives
	\begin{equation}
		X_{\mathrm{int}}(z)
		=
		\frac{1}{8\lambda\pi^2B^3}
		\left(
		\frac{1}{F}
		-
		\frac{1}{z}
		\right).
		\label{eq:Xint_closed_form_revised}
	\end{equation}
	which reveals that the tangent intercept is completely determined by the Airy beam parameters $(B,F)$ and the propagation distance $z$. Furthermore, differentiating \eqref{eq:Xint_closed_form_revised} with respect to $z$ yields
	\begin{equation}
		\frac{dX_{\mathrm{int}}(z)}{dz}
		=
		\frac{1}{8\lambda\pi^2B^3z^2}.
		\label{eq:Xint_monotonic_revised}
	\end{equation}
	Hence, $X_{\mathrm{int}}(z)$ is monotonic with respect to the propagation distance. As a result, the tightest one-sided aperture constraint is attained at the receiver plane $z=z_r$.
	
	To ensure physical realizability, the tangent intercept at $z_r$ must remain inside the finite transmit aperture. Therefore, the realizability condition is given by
	$X_{\mathrm{int}}(z_r)\le \frac{D_t}{2}, \quad B>0$,
	and
	$X_{\mathrm{int}}(z_r)\ge -\frac{D_t}{2}, \quad B<0$.
	Equivalently, a symmetric conservative condition can be written as
	\begin{equation}
		|X_{\mathrm{int}}(z_r)|
		\le
		\frac{D_t}{2}.
		\label{eq:realizability_condition}
	\end{equation}
	Substituting the closed-form solution of $F_\text{opt}$ \eqref{eq:closed_form_F} into \eqref{eq:Xint_closed_form_revised}, the tangent intercept at the receiver plane can be rewritten as
	\begin{equation}
		X_{\mathrm{int}}(z_r)
		=
		\frac{u}{v}
		+
		\frac{v}{16\lambda\pi^2B^3},
		\label{eq:Xint_uv_revised}
	\end{equation}
	where
	\begin{equation}
		u=\frac{x_s}{z_b}-\frac{x_r}{z_r},
		\qquad
		v=\frac{1}{z_b}-\frac{1}{z_r}.
		\label{eq:uv_revised}
	\end{equation}
	The above result serves as the foundation for deriving the effective geometric feasibility boundary of Airy beam generation under finite array apertures.
	
	\subsection{Geometric Feasible Boundary Induced by the Finite Aperture}
	\label{subsec:geometric_boundary}
	
	We next transform the tangent-intercept constraint into an explicit geometric feasibility boundary in the $(z_b,x_s)$ plane. The key observation is that based on the closed-form design in Sec.~\ref{sec:system_model}, the Airy beam parameters $\{B,F,\theta\}$ are not independent. Instead, for a given waypoint $(z_b,x_s)$ and receiver position $(z_r,x_r)$, these parameters are completely determined by the closed-form design in \eqref{eq:closed_form_B}--\eqref{eq:closed_form_theta}. Therefore, the physical realizability condition can be converted into a purely geometric feasibility condition. For notational simplicity, we define
	\begin{equation}
		X=\frac{x_s z_r-x_r z_b}
		{z_r-z_b}.
		\label{eq:X_definition_revised}
	\end{equation}
	By substituting the closed-form solution of $B_{\mathrm{opt}}$ \eqref{eq:closed_form_B} into \eqref{eq:Xint_uv_revised}, the tangent intercept can be rewritten as
	\begin{equation}
		X_{\mathrm{int}}(z_r)
		=
		X+\frac{1}{16\lambda\pi^2M(X)},
		\label{eq:Xint_scalar_revised_short}
	\end{equation}
	where
	\begin{equation}
		M(X)
		=
		\frac{3X}{16\lambda\pi^2\omega_0^2}
		+
		\sqrt{
			\left(
			\frac{3X}{16\lambda\pi^2\omega_0^2}
			\right)^2
			+
			\frac{3}{128\lambda^2\pi^4\omega_0^2}
		}.
		\label{eq:MX_appendix_ref}
	\end{equation}
	Equation \eqref{eq:Xint_scalar_revised_short} reveals that the finite-aperture feasibility condition collapses into a scalar equation solely determined by the geometric parameter $X$.
	
	For the case $B>0$, the upper feasible boundary is attained when the tangent intercept exactly reaches the upper edge of the transmit aperture, i.e., $X_{\mathrm{int}}(z_r)=\frac{D_t}{2}$.
	By solving the resulting scalar equation, the critical geometric parameter is obtained as
	\begin{equation}
		X^\star
		=
		\frac{5}{12}D_t.
		\label{eq:Xstar_short}
	\end{equation}
	
	The detailed derivation is provided in Appendix~\ref{appendix_boundary}.
	Then, using the definition of $X$ and setting $X=X^\star$, the feasible boundary 
	can be directly obtained.
	\begin{theorem}
		Under a finite aperture and $B>0$, the physically realizable waypoint region of Airy beam generation is bounded by
		\begin{equation}
			x_{s,\max}(z_b)
			=
			\frac{5D_t}{12}
			\left(
			1-\frac{z_b}{z_r}
			\right)
			+
			x_r\frac{z_b}{z_r}.
			\label{eq:xs_boundary_theorem}
		\end{equation}
		Any waypoint above this boundary requires a tangent intercept outside the physical transmit aperture and is therefore infeasible for Airy beam generation.
	\end{theorem}
	
	\begin{figure}[t]
		\centering
		\subfigure[Waypoint below the boundary.]{\includegraphics[width=0.8\linewidth]{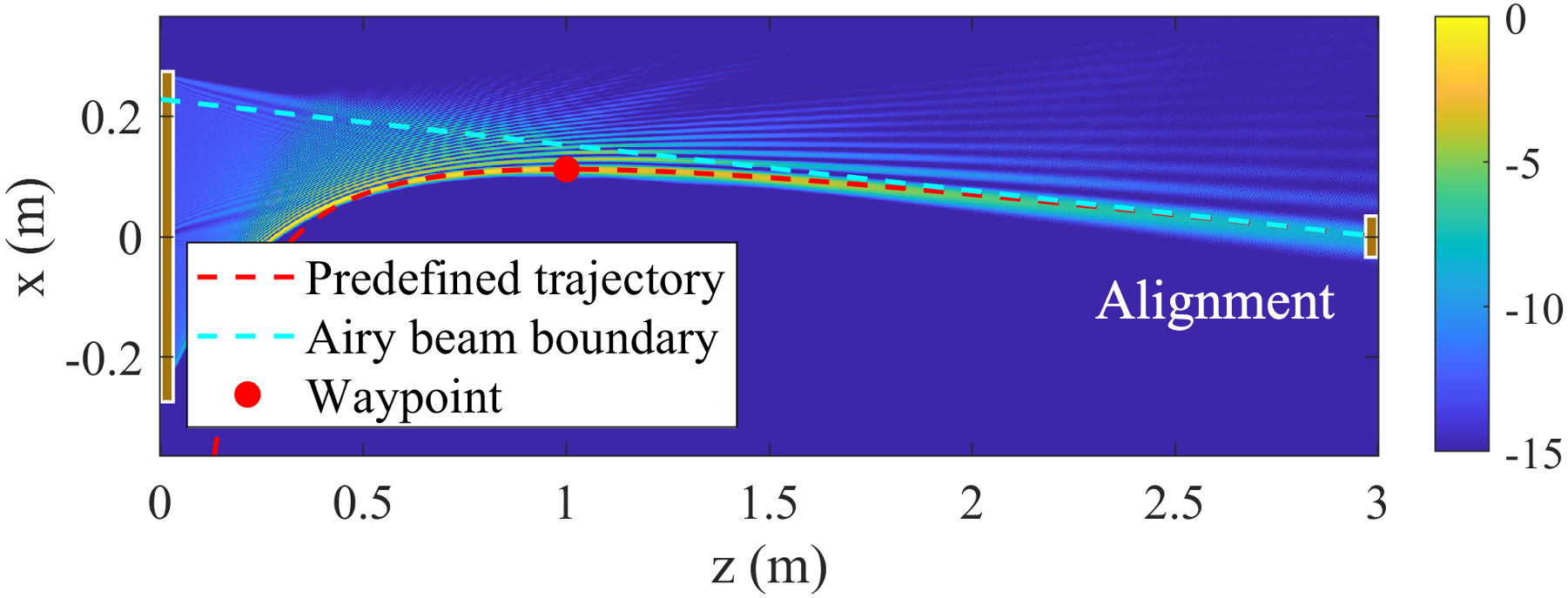}   \label{fig:bb1}}
		\subfigure[Waypoint on the boundary.]{\includegraphics[width=0.8\linewidth]{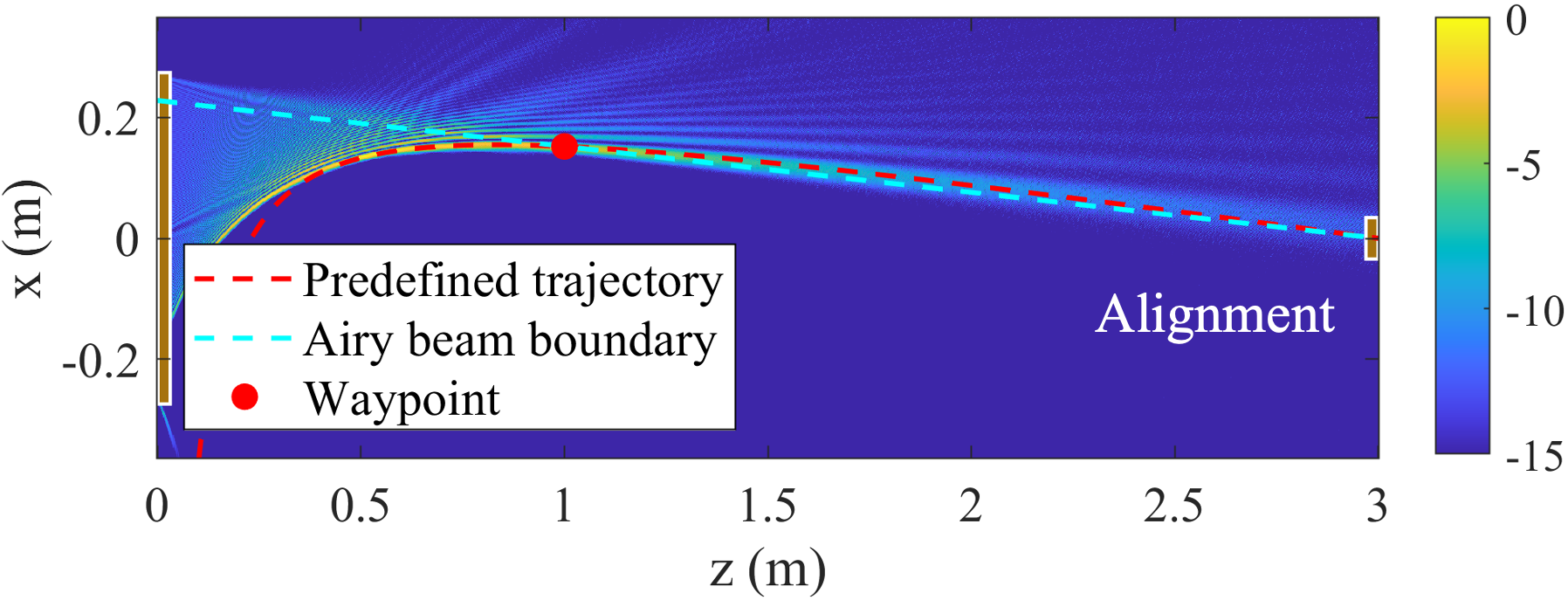}   \label{fig:bb2}}
		\subfigure[Waypoint above the boundary.]{\includegraphics[width=0.8\linewidth]{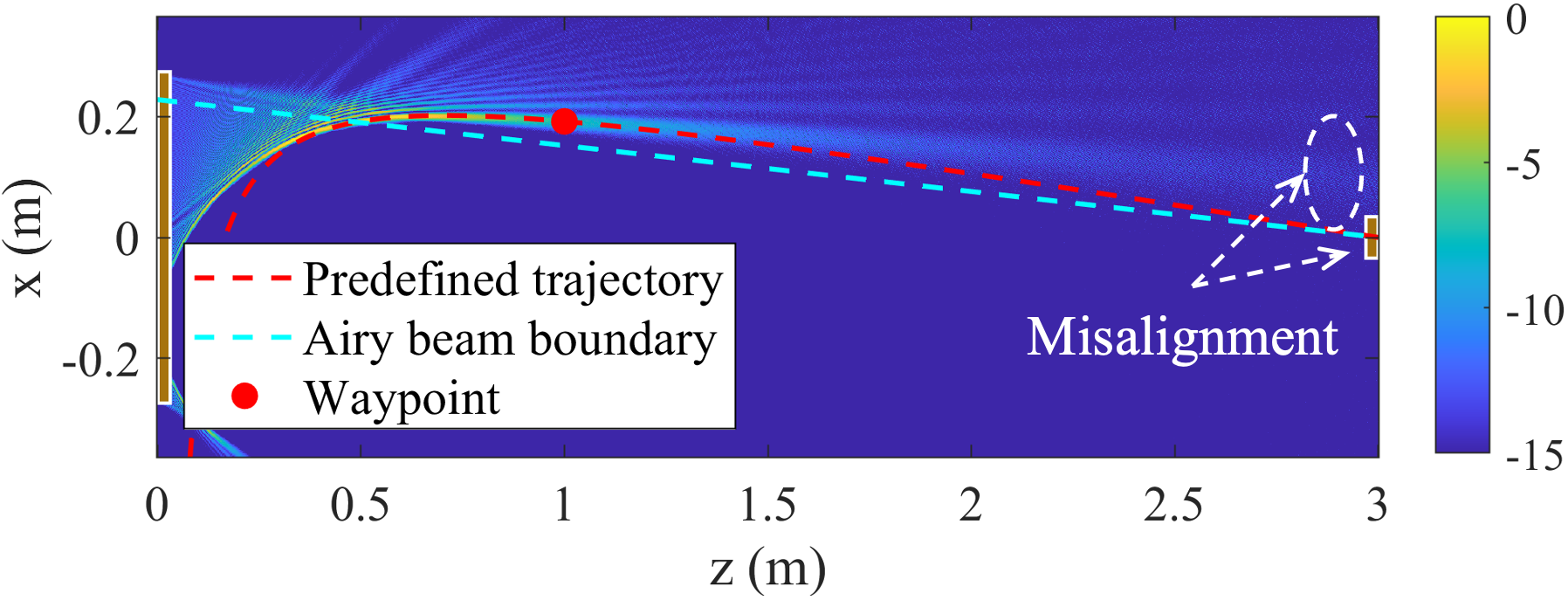}   \label{fig:bb3}}
		\caption{Airy beam boundary ($B>0$).}
		\label{fig:Airy_boundary}
	\end{figure}
	As shown in Fig.~\ref{fig:Airy_boundary}, the proposed finite-aperture feasibility boundary accurately characterizes the realizability limit of Airy beam generation for the case $B>0$. When the waypoint is located below or on the feasibility boundary, as illustrated in Fig.~\ref{fig:bb1} and Fig.~\ref{fig:bb2},  the tangent intercept remains inside the finite transmit aperture, such that the generated Airy beam can closely follow the predefined trajectory and align with the receiver. However, when the waypoint exceeds the feasibility boundary, as illustrated in Fig.~\ref{fig:bb3}, the tangent intercept falls outside the physical aperture, leading to aperture truncation, trajectory distortion, and receiver misalignment.
	
	By symmetry, for the case $B<0$, the tangent intercept is constrained by the lower edge of the transmit aperture, i.e., $X_{\mathrm{int}}(z_r)\ge -\frac{D_t}{2}$. Following a similar derivation, the corresponding lower feasible boundary can be obtained as
	\begin{equation}
		x_{s,\min}(z_b)
		=
		-\frac{5D_t}{12}
		\left(
		1-\frac{z_b}{z_r}
		\right)
		+
		x_r\frac{z_b}{z_r}.
		\label{eq:xs_boundary_lower}
	\end{equation}
	
	\section{Slope-Depth Geometric Codebook Design}
	\label{sec:non-uniform_codebook_design}
	
	In this section, we propose the SDGC, which aims to maximize the beam training performance while maintaining a manageable training overhead. Instead of sampling abstract Airy beam parameters, the proposed codebook directly samples candidate waypoints within the feasible LoS region. To this end, our design is divided into two sequential stages. First, we introduce an initial probing mechanism to determine the overall bending direction of the beam, thereby identifying the sign of $\sigma$ and effectively halving the required training overhead. Subsequently, motivated by the closed-form Airy beam design framework~\cite{Zhao-2026-Airy}, we develop a slope-depth geometric sampling strategy. The key intuition is that Airy beam trajectories are fundamentally governed by angular slope and inverse depth rather than Cartesian coordinates, resulting in a naturally non-uniform sampling distribution in the spatial domain. This strategy dynamically adapts the codebook density to the spatial propagation characteristics of Airy beams, thereby achieving a favorable trade-off between coverage resolution and training overhead.

	\subsection{Probing Mechanism}
	\label{susec:probing}
	
	Due to the inherent symmetry of the Airy function, an Airy beam can be designed to bend toward either the positive ($\sigma=+1$) or negative ($\sigma=-1$) transverse direction. In practical scenarios with unknown blockage locations, directly searching the codebook over both bending directions would approximately double the training overhead. To avoid this unnecessary search, we introduce an initial probing mechanism to determine the bending direction before beam training.
	
	\begin{figure}[t]
		\centering
		\includegraphics[width=0.8\linewidth]{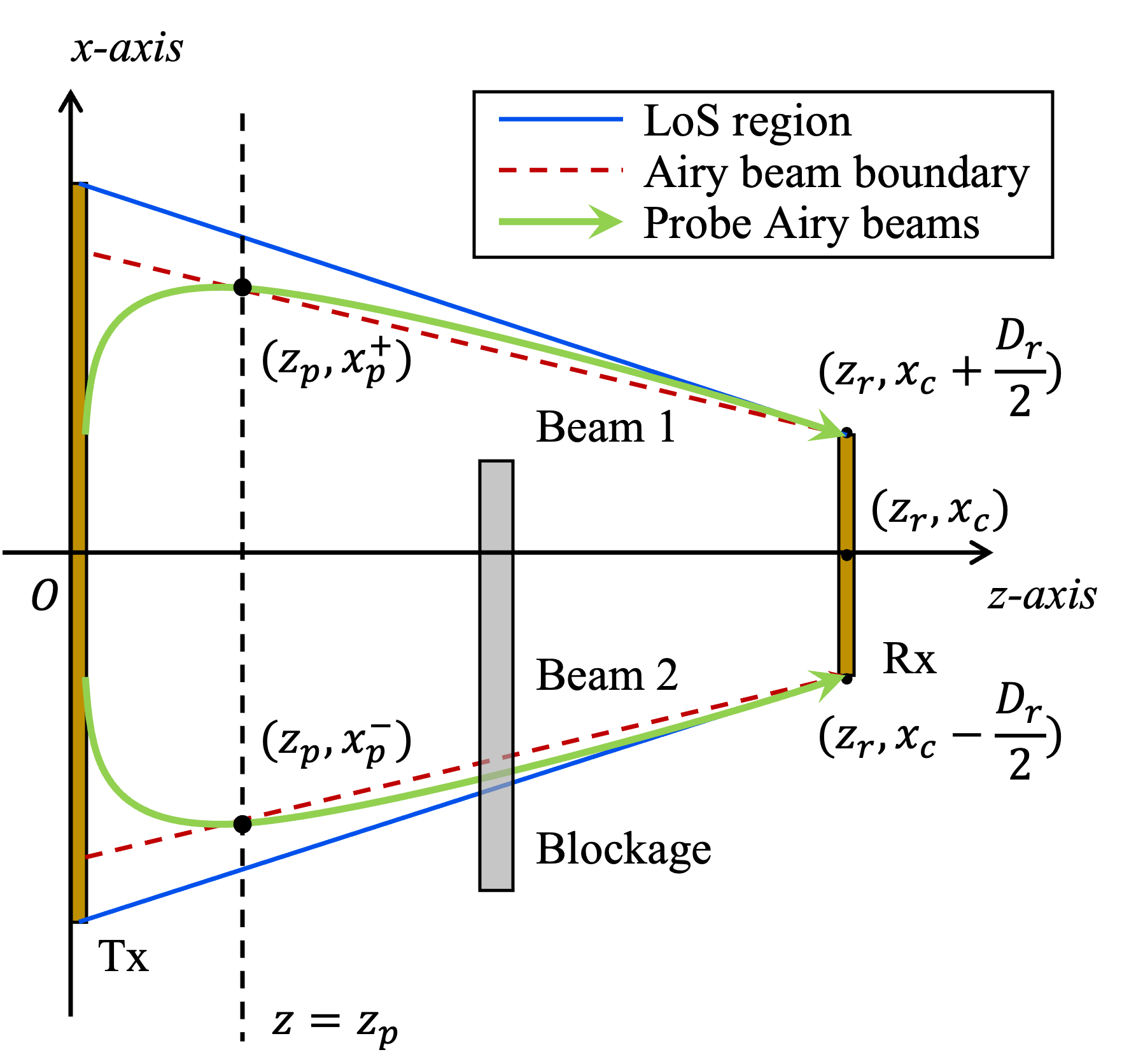}
		\caption{Probe beams design.}
		\label{fig:probe_beam}
	\end{figure}
	
	As shown in Fig.~\ref{fig:probe_beam}, we select a near-field reference plane at $z=z_p$, which intersects the theoretical Airy beam boundaries at the upper and lower alert points, denoted as $(z_p,x_p^+)$ and $(z_p,x_p^-)$, respectively. The two probe beams are designed to pass through these alert points and terminate at the corresponding edges of the receiver array, i.e., $(z_r,x_c+D_r/2)$ and $(z_r,x_c-D_r/2)$. Based on the closed-form Airy beam design in Sec.~\ref{sec:system_model}, their beamforming vectors are given by
	\begin{subequations}
		\begin{align}
			\mathbf{f}_p^{+}
			&=
			\mathcal{F}
			\left(
			\mathcal{A}
			\left(
			z_p,x_p^{+};
			z_r,x_c+\frac{D_r}{2}
			\right)
			\right),
			\\
			\mathbf{f}_p^{-}
			&=
			\mathcal{F}
			\left(
			\mathcal{A}
			\left(
			z_p,x_p^{-};
			z_r,x_c-\frac{D_r}{2}
			\right)
			\right).
		\end{align}
		\label{eq:probe_beams}
	\end{subequations}
	
	To determine the reference depth $z_p$, we establish the following lemma.
	
	\begin{lemma}
		\label{lemma:envelope}
		Let $S$ denote the spatial envelope area swept by an Airy probe beam between the alert point $(z_p,x_p)$ and the receiver target $(z_r,x_r)$. As $z_p$ approaches the transmitter, the envelope area increases and asymptotically approaches its physical maximum, thereby enlarging the spatial separation between the two probe beams.
	\end{lemma}
	
	\begin{IEEEproof}
		The proof is provided in Appendix~\ref{app:lemma_proof}.
	\end{IEEEproof}
	
	According to Lemma~\ref{lemma:envelope}, a smaller $z_p$ provides wider spatial coverage and larger separation between the two probe trajectories. However, decreasing $z_p$ also requires stronger Airy-beam bending, which redistributes more energy toward the side lobes and reduces the received main-lobe power. Therefore, $z_p$ is selected as the minimum reference depth that maintains sufficient received energy for reliable probing.
	
	Let $E^+$ and $E^-$ denote the received energies corresponding to $\mathbf{f}_p^+$ and $\mathbf{f}_p^-$, respectively. For a blockage extending upward from the lower side of the LoS region, the positive probe beam can bypass the blockage while the negative probe beam experiences stronger obstruction, resulting in $E^+>E^-$. Hence, $\sigma=+1$ is selected. Conversely, $E^- > E^+$ leads to $\sigma=-1$. Therefore, the bending direction is determined as
	\begin{equation}
		\sigma^\star
		=
		\begin{cases}
			+1, & E^+ \geq E^-,\\
			-1, & E^+ < E^-.
		\end{cases}
		\label{eq:probe_direction}
	\end{equation}
	This probing stage determines the bending direction before the subsequent SDGC or B1GC beam training and therefore avoids searching both bending branches.

	\subsection{Slope-Depth Geometric Codebook}
	
	Upon determining the bending direction $\sigma^\star$, SDGC is developed to perform beam training. According to the closed-form analytical framework in~\cite{Zhao-2026-Airy}, the optimal beam parameters $\{B,F,\theta\}$ can be uniquely derived once a spatial waypoint $(z_b,x_s)$ and the receiver target $(z_r,x_r)$ are specified. This closed-form solution identifies the specific Airy beam that circumvents a blockage at $(z_b,x_s)$ while maximizing the main-lobe power delivered to $(z_r,x_r)$. In scenarios where the exact blockage information is unavailable, we postulate a set of candidate waypoints as potential blockage points. By systematically sampling these waypoints, a comprehensive codebook is synthesized based on the closed-form solution to facilitate the subsequent training process. The optimal codeword is then selected based on the energy feedback from the receiver.
	
	Traditional beam training frameworks typically rely on uniform Cartesian sampling, where candidate waypoints are evenly distributed in the spatial domain. However, such a sampling strategy does not fully exploit the geometric structure of Airy beam formation and therefore leads to inefficient codebook construction. To develop a geometry-based sampling framework, we draw mathematical inspiration directly from the derived closed-form solution of the optimal bending parameter $B_{\rm opt}$. A careful inspection of \eqref{eq:closed_form_B} reveals that the required beam curvature is primarily governed by two fundamental geometric quantities, namely the angular slope $x_s/z_b$ and the inverse depth $1/z_b$. Motivated by this observation, we introduce a coordinate transformation from the conventional Cartesian space $(x,z)$ to a slope-depth domain parameterized by the angular slope $\phi$ and inverse depth $\gamma$, where
	\begin{equation}
		\phi=\frac{x}{z},
		\qquad
		\gamma=\frac{1}{z}.
		\label{eq:slope_depth_transform}
	\end{equation}
	By directly sampling in the $(\phi,\gamma)$ domain, the codebook construction naturally aligns with the intrinsic geometry of Airy beam formation, enabling more efficient beam training than conventional Cartesian-domain sampling.
	
	Specifically, we first define the search interval for the depth $z_b$ as $[z_{\min},z_{\max}]$, which corresponds to the inverse-depth range $\gamma\in[\gamma_{\min},\gamma_{\max}]$ with $\gamma=1/z_b$. To determine the angular-slope range, we consider the LoS region boundaries, which are defined as the linear envelopes connecting the edges of the transmitter and receiver apertures. The upper LoS boundary $x_{\mathrm{up}}$ at $z_{\min}$ is expressed as
	\begin{equation}
		x_{\mathrm{up}}
		=
		\frac{D_r-D_t}{2z_r}z_{\min}
		+
		\frac{D_t}{2}.
	\end{equation}
	Consequently, the angular-slope range $[-\phi_{\max},\phi_{\max}]$ is determined by the geometric intersections of these boundaries, where
	\begin{equation}
		\phi_{\max}
		=
		\frac{x_{\mathrm{up}}}{z_{\min}}.
	\end{equation}
	
	Within this transformed rectangular domain $(\gamma,\phi)$, we perform uniform sampling to construct the SDGC. The sampling resolutions are determined by the prescribed quantization levels $M$ and $N$, which control the tradeoff between beam training resolution and training overhead. Specifically,
	\begin{equation}
		\Delta\gamma
		=
		\frac{\gamma_{\max}-\gamma_{\min}}{M-1},
		\qquad
		\Delta\phi
		=
		\frac{2\phi_{\max}}{N-1}.
		\label{eq:SDGC_stepsize}
	\end{equation}
	Accordingly, for a given beam-training budget, $M$ and $N$ can be selected to adjust the sampling resolutions along the inverse-depth and angular-slope dimensions, respectively. The resulting sampling points are given by
	\begin{equation}
		\begin{cases}
			\gamma_m
			=
			\gamma_{\min}
			+
			m\Delta\gamma,
			&
			m=0,1,\dots,M-1,
			\\
			\phi_n
			=
			-\phi_{\max}
			+
			n\Delta\phi,
			&
			n=0,1,\dots,N-1.
		\end{cases}
		\label{eq:SDGC_sampling}
	\end{equation}
	Before feasibility pruning, the transformed grid therefore contains $K=MN$ candidate waypoints.
	
	For each linear index $i\in\{1,2,\dots,K\}$, the corresponding grid indices $(m,n)$ are uniquely determined via
	\begin{equation}
		n=(i-1)\pmod N,
		\qquad
		m=\lfloor(i-1)/N\rfloor.
	\end{equation}
	By applying the inverse coordinate transformation, the $i$-th Cartesian waypoint $(z_b^{(i)},x_s^{(i)})$ used to synthesize the Airy codeword is explicitly given by
	\begin{equation}
		\begin{cases}
			z_b^{(i)}
			=
			\displaystyle\frac{1}{\gamma_m}
			=
			\displaystyle\frac{1}{\gamma_{\min}+m\Delta\gamma},
			\\[3mm]
			x_s^{(i)}
			=
			\displaystyle\frac{\phi_n}{\gamma_m}
			=
			\displaystyle\frac{-\phi_{\max}+n\Delta\phi}
			{\gamma_{\min}+m\Delta\gamma}.
		\end{cases}
		\label{eq:SDGC_inverse_transform}
	\end{equation}
	
	\begin{figure}[t]
		\centering
		\includegraphics[width=0.8\linewidth]{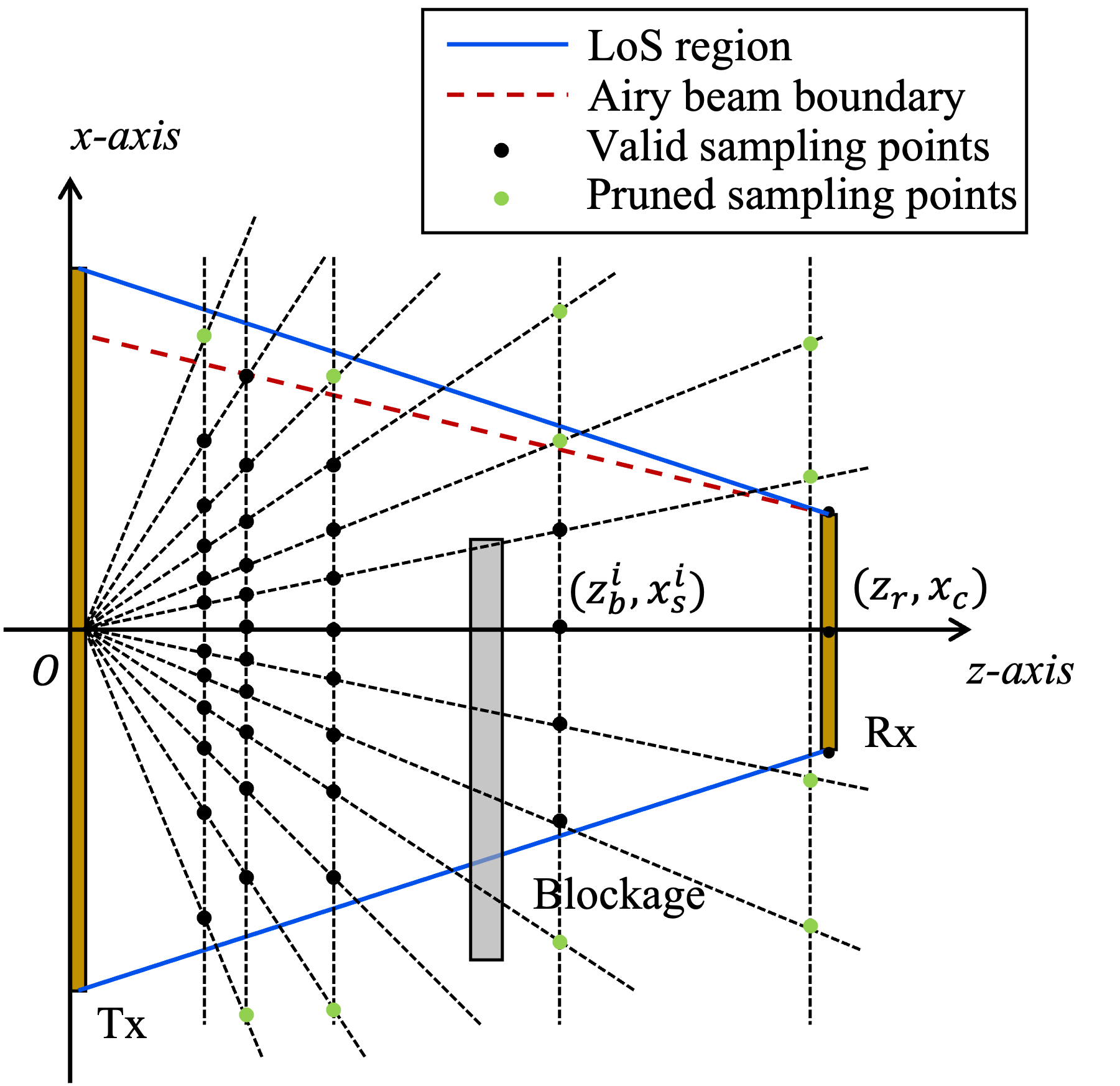}
		\caption{Slope-depth geometric codebook design.}
		\label{fig:NUGC_design}
	\end{figure}
	
	Taking $\sigma^\star=+1$ as an example, the distribution of the proposed SDGC is illustrated in Fig.~\ref{fig:NUGC_design}. Specifically, uniform sampling in the slope-depth domain results in a non-uniform distribution of candidate waypoints in the Cartesian spatial domain rather than a conventional uniform Cartesian grid. Since the blockage only appears within the LoS region, we retain only the waypoints located both in the LoS region and below the Airy beam boundary to ensure effective blockage evasion. Consequently, waypoints that fall outside this valid region are pruned. This pruning strategy focuses the synthesized codebook on physically valid blockage-evasion scenarios, thereby reducing the codebook size and beam training overhead.
	
	For each retained candidate waypoint $(z_b^{(i)},x_s^{(i)})$, a corresponding receiver focal point $(z_r,x_r^{(i)})$ must be determined to ensure that the waypoint lies within the traversable region of the Airy beam. According to the boundary equation in \eqref{eq:xs_boundary_theorem}, the lower bound of $x_r^{(i)}$ that guarantees the waypoint $(z_b^{(i)},x_s^{(i)})$ remains below the Airy beam boundary is expressed as
	\begin{equation}
		x_{\mathrm{bound}}^{(i)}
		=
		\frac{z_r}{z_b^{(i)}}x_s^{(i)}
		-
		\frac{5D_t}{12}
		\left(
		\frac{z_r}{z_b^{(i)}}-1
		\right).
	\end{equation}
	
	From an energy perspective, $x_r^{(i)}$ should be minimized to maximize the main-lobe power delivered to the receiver aperture. Consequently, to ensure successful blockage evasion while maintaining efficient energy reception, $x_r^{(i)}$ is selected as the smallest possible value satisfying the boundary constraint, without falling below the center of the Rx antenna array. Furthermore, considering the physical size of the Rx, $x_r^{(i)}$ is upper-bounded by $x_c+D_r/2$ to ensure that the beam remains directed toward the effective aperture. Therefore, the receiver target is given by
	\begin{equation}
		x_r^{(i)}
		=
		\min
		\left(
		x_c+\frac{D_r}{2},
		\max
		\left(
		x_c,
		x_{\mathrm{bound}}^{(i)}
		\right)
		\right).
		\label{eq:SDGC_rx_target}
	\end{equation}
	
	Finally, the corresponding Airy codeword is constructed by substituting the retained waypoint and receiver target into the closed-form design operator as
	\begin{equation}
		\mathbf{f}^{(i)}
		=
		\mathcal{F}
		\left(
		\mathcal{A}
		\left(
		z_b^{(i)},
		x_s^{(i)};
		z_r,
		x_r^{(i)}
		\right)
		\right).
		\label{eq:SDGC_codeword}
	\end{equation}
	A similar procedure applies to the case of $\sigma^\star=-1$. The proposed SDGC aligns the codebook sampling with the geometric characteristics of Airy beam propagation and the closed-form design framework, thereby increasing the likelihood of identifying an effective blockage-avoiding trajectory with reduced beam training overhead.

	\begin{figure}[t]
		\centering
		\includegraphics[width=0.8\linewidth]{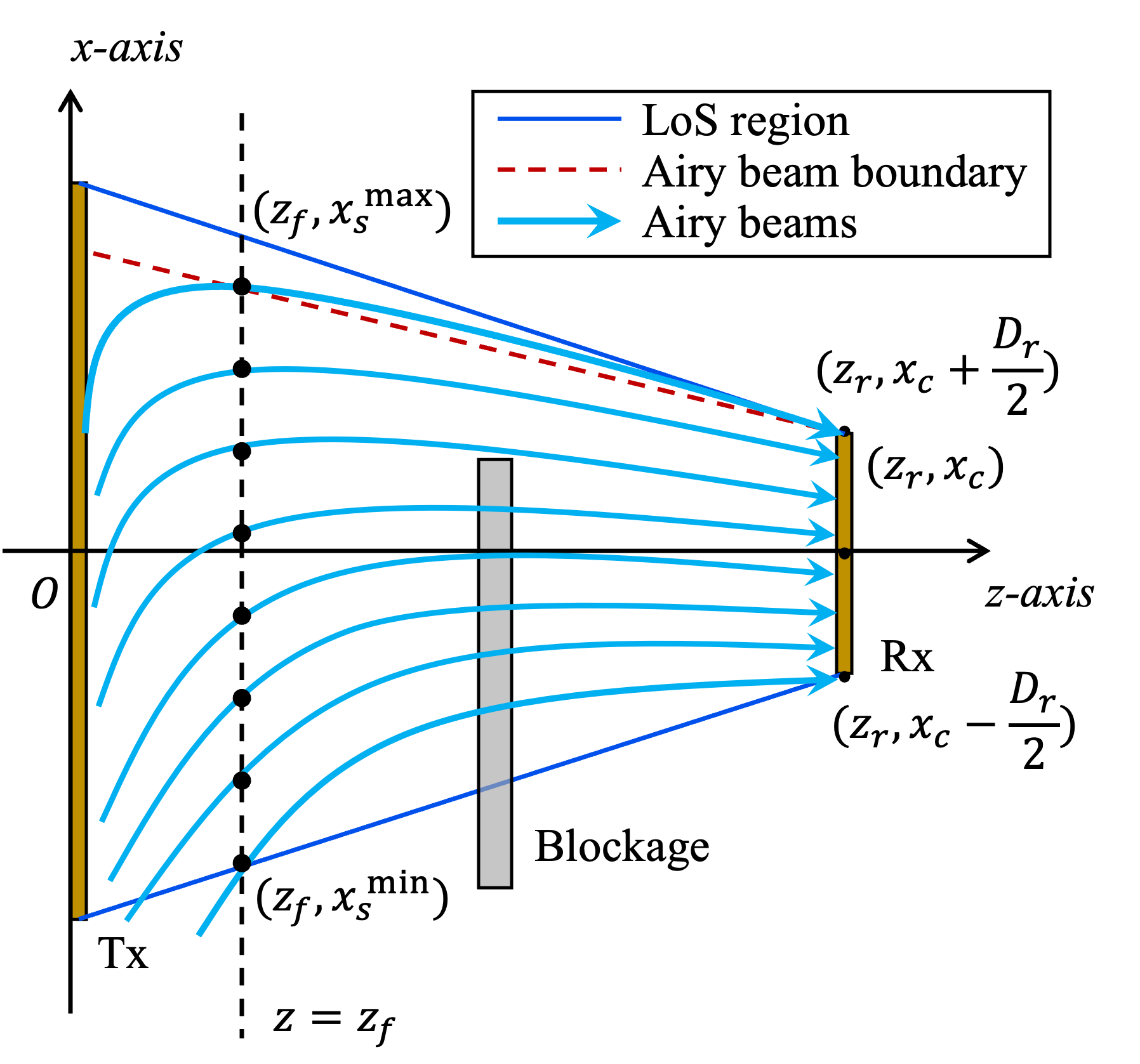}
		\caption{Bijective 1D geometric codebook design.}
		\label{fig:FS1C}
	\end{figure}
	
	\section{Bijective 1D Geometric Codebook Design}
	\label{sec:fast scanning}
	Although SDGC substantially reduces the beam training overhead, it still requires a two-dimensional search over the candidate waypoint coordinates $(z_b,x_s)$. To further reduce the training latency in dynamic THz environments, where blockages may change rapidly, we develop the B1GC as a low-dimensional beam training scheme for rapid link establishment. 
	
	The key idea originates from the unique propagation characteristics of Airy beams. Unlike focusing beams that concentrate energy at a single spatial point, Airy beams exhibit self-acceleration and non-diffraction, resulting in an extended energy distribution along a curved trajectory. Consequently, each Airy beam naturally covers a one-dimensional spatial manifold within the LoS region. This property implies that a properly designed sequence of Airy trajectories can sweep across the LoS region using significantly fewer codewords than conventional point-focusing beam training schemes. By exploiting this extended spatial coverage, the proposed B1GC enables efficient geometry-based scanning while substantially reducing the beam training latency.
	
	Following the probing mechanism in Sec.~\ref{susec:probing}, the bending direction $\sigma^\star$ is first determined by comparing the received energies of the two initial probe beams. The longitudinal coordinate of the waypoint is then fixed at the near-field reference plane $z_f=z_{\min}$ according to Lemma~\ref{lemma:envelope}, reducing the original 2D search space $(z_b,x_s)$ to a 1D search over $x_s$. This design places the waypoint close to the transmitter, thereby enlarging the spatial spread of the resulting Airy trajectories and increasing the achievable spatial coverage.
	
	To construct a deterministic 1D sweep, we establish a bijective linear mapping between the near-field waypoint $x_s$ and the target focal point $x_r$ on the receiver array. Specifically, let the waypoint scanning interval be $x_s\in[x_{s,\min},x_{s,\max}]$, and the corresponding reception range be $x_r\in[x_c-D_r/2,x_c+D_r/2]$. The target point $x_r$ is linked to $x_s$ through
	\begin{equation}
		\label{eq:1D_mapping}
		x_r(x_s)
		=
		\left(
		x_c-\frac{D_r}{2}
		\right)
		+
		\frac{x_s-x_{s,\min}}
		{x_{s,\max}-x_{s,\min}}
		D_r.
	\end{equation}
	This mapping establishes a one-to-one correspondence between the candidate waypoint and the receiver target over their respective intervals. Substituting \eqref{eq:1D_mapping} into the closed-form Airy beam design equations, the entire beamforming profile becomes fully parameterized by the single variable $x_s$.
	
	The primary concern of the proposed 1D scanning strategy is whether the reduced-dimensional search introduces spatial blind spots within the LoS region. To address this issue, we establish the following spatial coverage property of the resulting Airy trajectory family.
	
	\begin{lemma}
		\label{lemma:coverage}
		By continuously sweeping the waypoint parameter $x_s$ within its feasible interval, the resulting family of Airy beam trajectories continuously covers the spatial region bounded by the two trajectories corresponding to $x_{s,\min}$ and $x_{s,\max}$.
	\end{lemma}
	
	\begin{IEEEproof}
		The proof is provided in Appendix~\ref{app:coverage_proof}.
	\end{IEEEproof}
	
	\textit{Lemma}~\ref{lemma:coverage} establishes the continuous spatial coverage of the trajectory family between the two boundary trajectories. For practical beam training, the continuous interval $[x_{s,\min},x_{s,\max}]$ is uniformly discretized into $P$ candidate waypoints. The corresponding sampling interval is therefore determined as
	\begin{equation}
		\Delta x_s
		=
		\frac{x_{s,\max}-x_{s,\min}}{P-1},
		\label{eq:B1GC_stepsize}
	\end{equation}
	and the resulting discrete waypoint set is
	\begin{equation}
		x_s^{(p)}
		=
		x_{s,\min}
		+
		p\Delta x_s,
		\qquad
		p=0,1,\dots,P-1.
		\label{eq:B1GC_waypoints}
	\end{equation}
	Thus, increasing $P$ provides a finer approximation of the continuous trajectory family at the cost of additional beam training overhead.
	
	For each waypoint $x_s^{(p)}$, the corresponding receiver target $x_r^{(p)}$ is obtained from the bijective mapping in \eqref{eq:1D_mapping} as
	\begin{equation}
		x_r^{(p)}
		=
		x_r\left(x_s^{(p)}\right).
	\end{equation}
	The corresponding Airy codeword is then generated through the closed-form design operator as
	\begin{equation}
		\mathbf{f}^{(p)}
		=
		\mathcal{F}
		\left(
		\mathcal{A}
		\left(
		z_f,
		x_s^{(p)};
		z_r,
		x_r^{(p)}
		\right)
		\right).
		\label{eq:B1GC_codeword}
	\end{equation}
	Accordingly, the B1GC codebook is expressed as
	\begin{equation}
		\mathcal{C}_{\mathrm{B1GC}}
		=
		\left\{
		\mathbf{f}^{(p)}
		\right\}_{p=0}^{P-1}.
	\end{equation}
	
	During beam training, the BS sequentially transmits the codewords in $\mathcal{C}_{\mathrm{B1GC}}$, and the codeword yielding the largest received signal power is selected. Compared with exhaustive 2D beam training, B1GC reduces the search complexity from $\mathcal{O}(MN)$ to $\mathcal{O}(P)$ and therefore significantly lowers the training latency. Since the proposed 1D search prioritizes spatial coverage rather than exhaustive coordinate optimization, the selected beam does not necessarily maximize the received energy over the entire 2D waypoint space. Nevertheless, B1GC enables rapid identification of an effective Airy propagation trajectory with substantially reduced training overhead, making it particularly suitable for highly dynamic THz blockage scenarios.

	\section{Geometry-Based Airy Beam Training for UPA Systems}
	\label{sec:UPA}
	
	Based on the hybrid focusing--Airy mode introduced in our previous work~\cite{Zhao-2026-Airy}, the probing mechanism in Sec.~\ref{susec:probing} can be extended to the two transverse dimensions of a UPA. Specifically, the original probe beams $\mathbf{f}_p^{+}$ and $\mathbf{f}_p^{-}$ are constructed along both the $x$- and $y$-dimensions, resulting in four hybrid probe beams
	\begin{equation}
		\begin{aligned}
			\mathbf{f}_{p,x}^{\pm}
			&=
			\mathbf{f}_{p,x}^{\mathrm{A},\pm}
			\otimes
			\mathbf{f}_{y}^{\mathrm{F}},
			\\
			\mathbf{f}_{p,y}^{\pm}
			&=
			\mathbf{f}_{x}^{\mathrm{F}}
			\otimes
			\mathbf{f}_{p,y}^{\mathrm{A},\pm},
		\end{aligned}
		\label{eq:upa_probe}
	\end{equation}
	where $\mathbf{f}_{p,\nu}^{\mathrm{A},\pm}$ denotes the Airy probe beam bending toward the positive or negative direction along transverse dimension $\nu\in\{x,y\}$, while $\mathbf{f}_{x}^{\mathrm{F}}$ and $\mathbf{f}_{y}^{\mathrm{F}}$ denote the near-field focusing vectors along the $x$- and $y$-dimensions, respectively.
	
	Let $E_{\nu}^{+}$ and $E_{\nu}^{-}$ denote the received energies corresponding to the positive and negative probe beams along dimension $\nu$. The transverse dimension for blockage avoidance and the corresponding bending direction are jointly determined as
	\begin{equation}
		(\nu^\star,\sigma^\star)
		=
		\arg\max_{\nu\in\{x,y\},\,\sigma\in\{+1,-1\}}
		E_{\nu}^{\sigma},
		\label{eq:upa_probe_selection}
	\end{equation}
	where $\nu^\star$ specifies the transverse dimension used for blockage avoidance and $\sigma^\star$ determines the corresponding bending direction.
	
	After the probing stage, the subsequent beam training is performed only along the selected dimension $\nu^\star$, while near-field focusing is retained along the orthogonal dimension. Therefore, the proposed SDGC and B1GC can be directly applied along the selected transverse dimension without jointly searching the two-dimensional UPA beam space. Specifically, the $k$-th UPA beamforming vector is expressed as
	\begin{equation}
		\mathbf{f}_{k}^{\mathrm{UPA}}
		=
		\begin{cases}
			\mathbf{f}_{x,k}^{\mathrm{A}}(\sigma^\star)
			\otimes
			\mathbf{f}_{y}^{\mathrm{F}},
			& \nu^\star=x,
			\\[1mm]
			\mathbf{f}_{x}^{\mathrm{F}}
			\otimes
			\mathbf{f}_{y,k}^{\mathrm{A}}(\sigma^\star),
			& \nu^\star=y,
		\end{cases}
		\label{eq:upa_training}
	\end{equation}
	where $\mathbf{f}_{\nu,k}^{\mathrm{A}}(\sigma^\star)$ denotes the $k$-th Airy codeword generated by SDGC or B1GC along the selected dimension. In this manner, the proposed framework extends the ULA beam training schemes to UPA systems with only four initial probe beams, while avoiding joint beam training over both transverse dimensions.
	\section{Simulation Results}
	\label{sec:simulation_results}
	
	In this section, simulations are conducted to evaluate the performance of the proposed Airy beam training codebooks under quasi-LoS THz near-field communication scenarios. The carrier frequency is set to 140~GHz with a bandwidth of 1~GHz~\cite{Han-2015-Multi-Ray}. For the ULA system, the transmitter and receiver are equipped with $N_t=512$ and $N_r=256$ antenna elements, respectively, and the communication distance is fixed at $D=3$~m. A blockage is located within the LoS region, resulting in a quasi-LoS propagation environment in which the direct transmission path is partially obstructed.
	
	The probing plane used to determine the bending direction is located at $z_p=0.5$~m. For the proposed SDGC, the inverse-depth parameter is sampled over $\gamma\in[1/3,\,2]$ with $M=5$ quantization levels, corresponding to a sampling interval of
	$\Delta\gamma\approx0.417$~m$^{-1}$. The angular-slope parameter is sampled over $\phi\in[-0.5018,\,0.5018]$ with $N=40$ quantization levels, corresponding to $\Delta\phi\approx0.0257$.
	
	For the proposed B1GC, the reference depth is set to $z_f=0.5$~m. The waypoint scanning range is configured as $x_s\in[-0.2509,\,0.1901]$~m and uniformly discretized into $P=20$ candidate waypoints, corresponding to a sampling interval of $\Delta x_s\approx0.0232$~m. The values of $M$, $N$, and $P$ are selected to provide a practical balance between spatial sampling resolution and beam training overhead.

	To assess the proposed Airy beam codebooks, we select the following benchmarks for comparison:
	
	\begin{itemize}
		
		\item \textbf{Quasi-LoS full-digital precoding (QL-FDP):} Assuming perfect knowledge of the quasi-LoS channel $\mathbf{H}_b$, the full-digital precoder is designed using Maximum Ratio Transmission (MRT) and Maximum Ratio Combining (MRC) to provide a single-stream performance upper bound on spectral efficiency in the quasi-LoS scenario.
		
		\item \textbf{LoS full-digital precoding (LoS-FDP):} The full-digital precoder is designed using MRT and MRC based on the LoS channel, while the spectral efficiency is evaluated under the quasi-LoS scenario.
		
		\item \textbf{Far-field steering beam search (FF-SBS)}~\cite{Alkhateeb-2015-Limited}: This scheme employs a conventional DFT codebook to search for the optimal far-field steering beam. The steering angle is uniformly sampled over the range $\sin\theta\in[-1,1]$ with a sampling interval of $s_\theta=0.056$.
		
		\item \textbf{Near-field focusing beam search (NF-FBS)}~\cite{Zhang-2022-Beam-focusing}: This scheme adopts the conventional polar-domain codebook to identify the optimal near-field focusing beam. The distance parameter is sampled in the polar domain according to $\cos^2\theta/r\in[1/3,\,7/3]$ with a sampling interval of $s_r=1/3$. The angle sampling strategy is identical to that of FF-SBS.
		
		\item \textbf{Hierarchical focusing--Airy codebook (HFAC)}~\cite{Zhao-2025-WDC}: This scheme performs hierarchical Airy beam training, where the optimal focusing distance-angle pair is first determined, followed by a refinement stage to identify the optimal curving parameter. The sampling of the distance-angle pair follows the same configuration as NF-FBS, while the curving parameter is sampled over the range $a\in[-3,3]$ with a sampling interval of $s_a=0.25$.
		
		\item \textbf{Exhaustive Airy beam search (EABS)}~\cite{Chen-2024-Curving}: This scheme exhaustively searches over the Airy beam parameter space to identify the parameter combination $\{B,F,\theta\}$ that maximizes the spectral efficiency. Specifically, the bending parameter is sampled over $B\in[-4,4]$ using 60 uniformly spaced samples, the focal distance is sampled over $F\in[0.5,3]$ using 30 samples, and the steering angle is sampled over $\theta\in[-\pi/4,\pi/4]$ using 40 samples.
		
	\end{itemize}
	
	\begin{figure}[t]
		\centering
		\subfigure[LoS-FDP.]{%
			\includegraphics[width=0.8\linewidth]{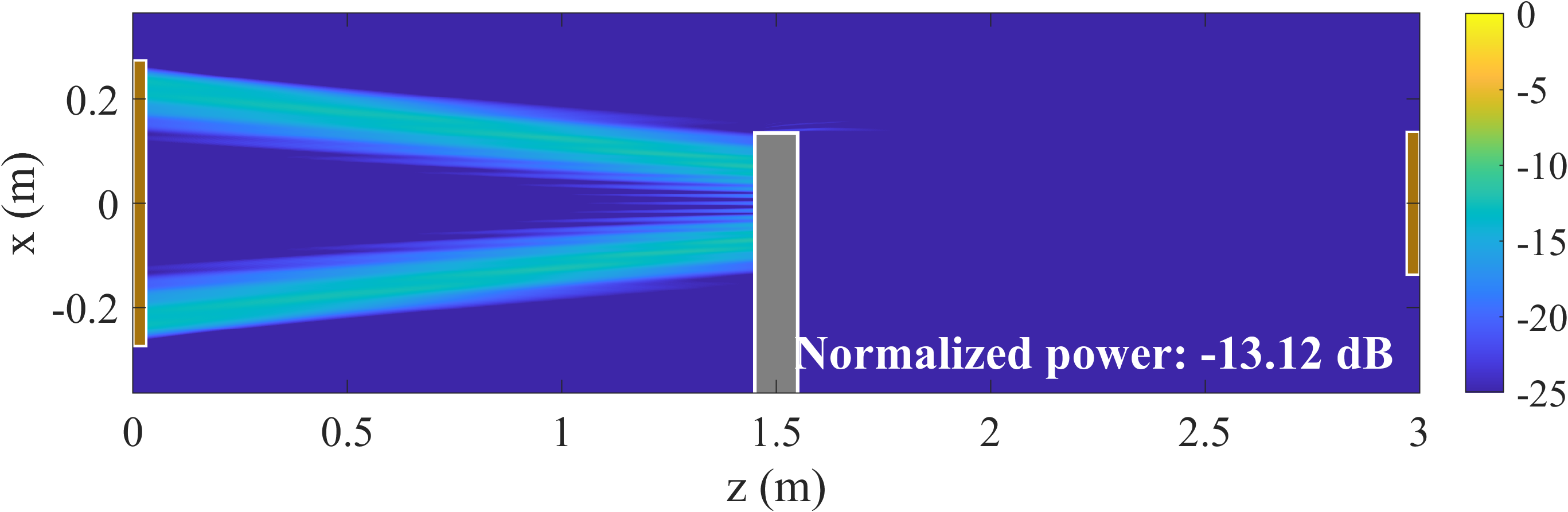}
			\label{fig:LoS_precoder}
		}  
		\subfigure[SDGC.]{%
			\includegraphics[width=0.8\linewidth]{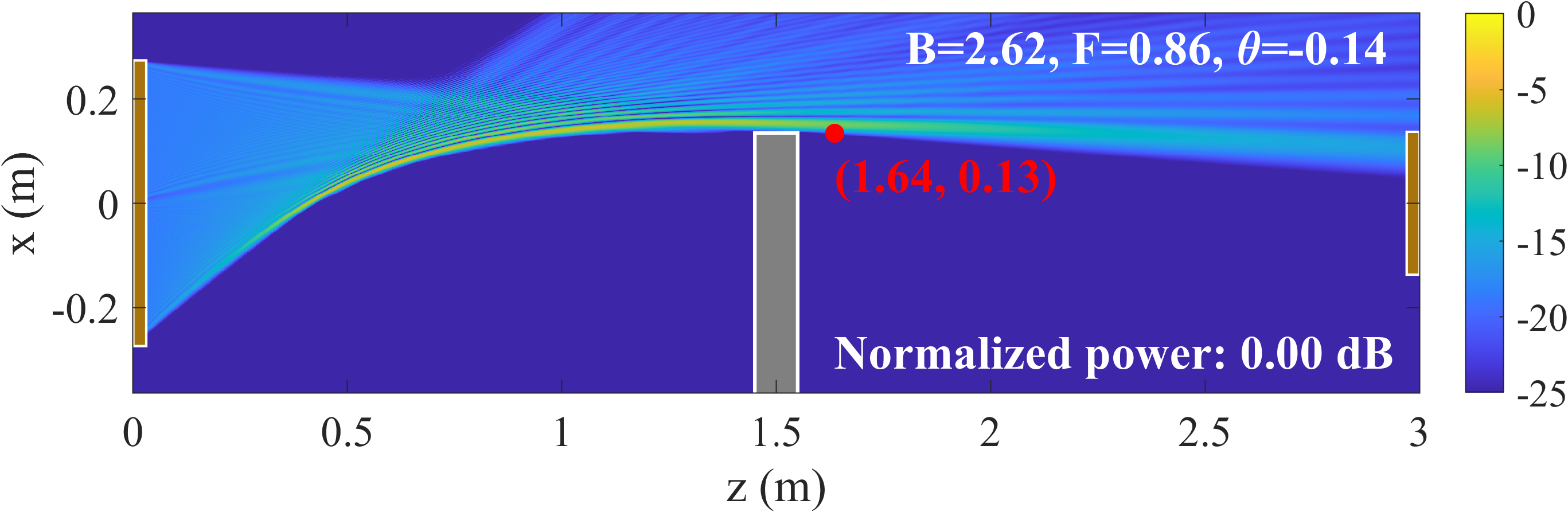}
			\label{fig:NUGC}
		}
		\subfigure[B1GC.]{%
			\includegraphics[width=0.8\linewidth]{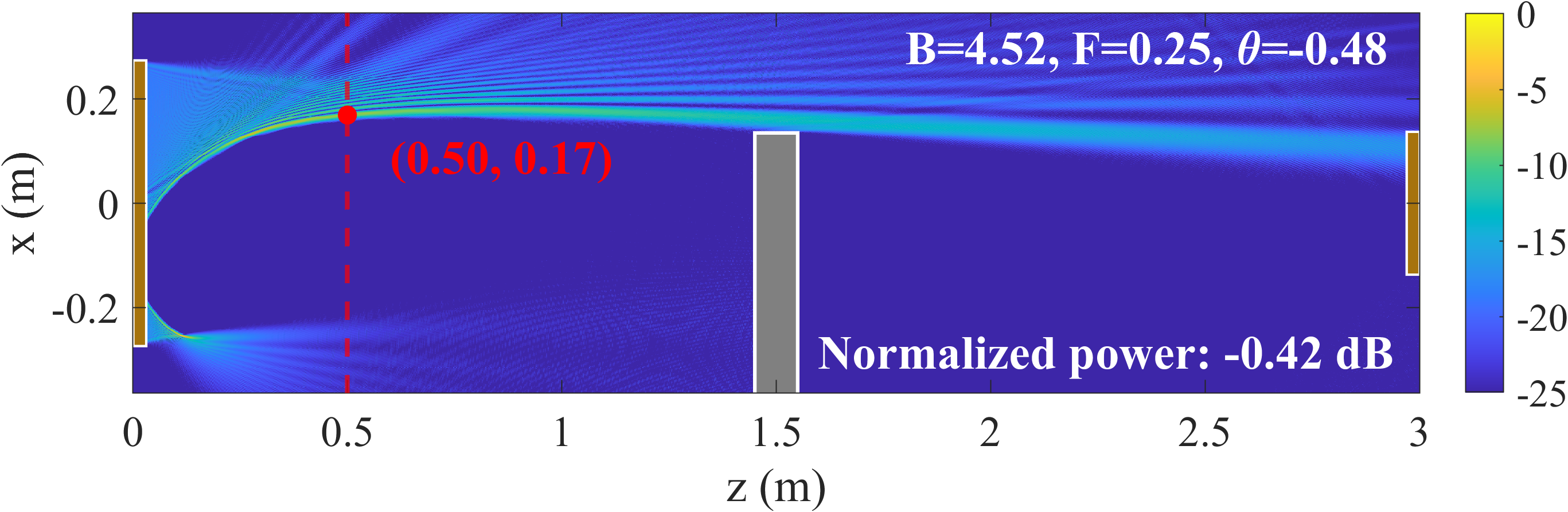}
			\label{fig:FS1C_beam}
		}
		\subfigure[HFAC.]{%
			\includegraphics[width=0.8\linewidth]{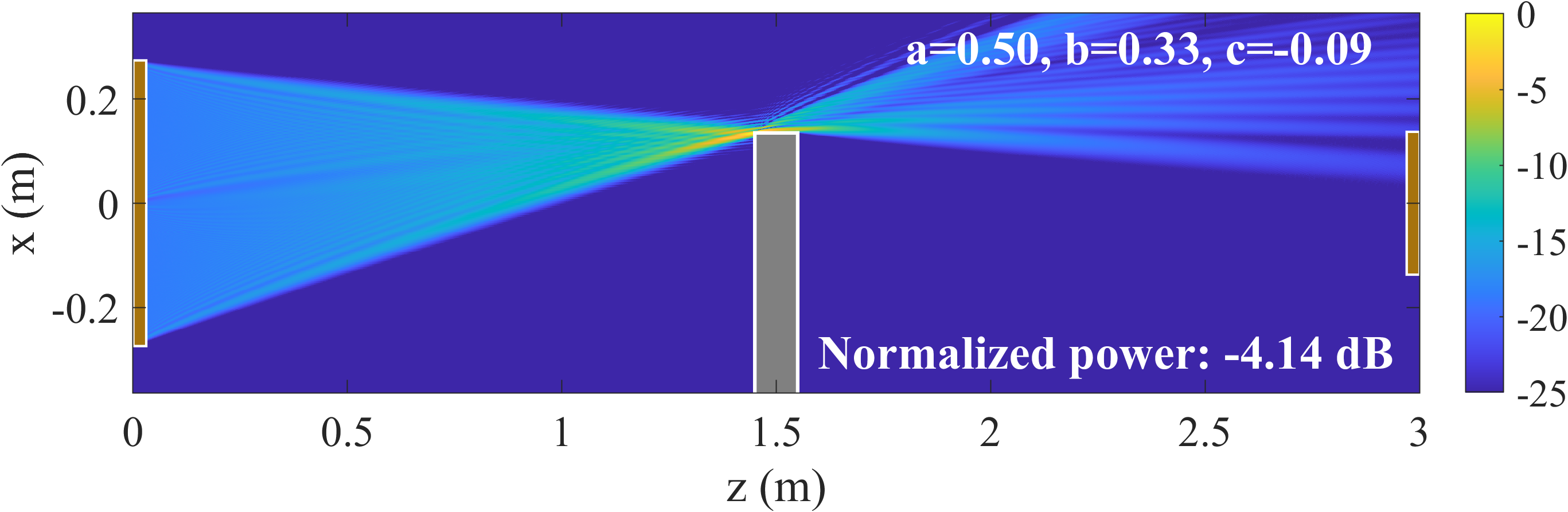}
			\label{fig:HFAC}
		}
		\caption{Beam propagation comparison among LoS-FDP, the proposed SDGC and B1GC, and HFAC under a quasi-LoS scenario with a 65.8\% blockage ratio.}
		\label{fig:beam_compare}
	\end{figure}

	Fig.~\ref{fig:beam_compare} illustrates the beam propagation patterns of different beamforming strategies under a quasi-LoS scenario. A blockage with height $0.135$~m is placed at $z=1.5$~m, resulting in a blockage ratio of $65.8\%$. The received power is normalized with respect to the best-performing beam.
	As shown in Fig.~\ref{fig:LoS_precoder}, the LoS-FDP beam propagates directly along the LoS path and is severely blocked by the obstacle, resulting in a normalized received-power loss of $13.12$~dB. In contrast, the proposed SDGC in Fig.~\ref{fig:NUGC} identifies an effective waypoint at $(1.64,0.13)$ and enables the Airy beam to bypass the blockage while maintaining accurate alignment with the receiver. SDGC achieves the highest received power, providing gains of $13.12$~dB and $4.14$~dB over LoS-FDP and HFAC, respectively.
	Fig.~\ref{fig:FS1C_beam} further demonstrates the proposed B1GC. By performing a 1D spatial sweep at the reference depth $z_f$, B1GC identifies a feasible trajectory through the waypoint $(0.50,0.17)$. Although B1GC prioritizes low training overhead rather than peak received power, it incurs only a $0.42$~dB loss compared with SDGC. For comparison, HFAC in Fig.~\ref{fig:HFAC} suffers a $4.14$~dB loss relative to SDGC. This is because HFAC performs hierarchical search over the focusing
	distance--angle pair and the curving parameter in the beam-parameter
	domain, rather than directly adapting the Airy trajectory to the
	blockage geometry.

	\begin{figure}[t]
		\centering
		\includegraphics[width=0.8\linewidth]{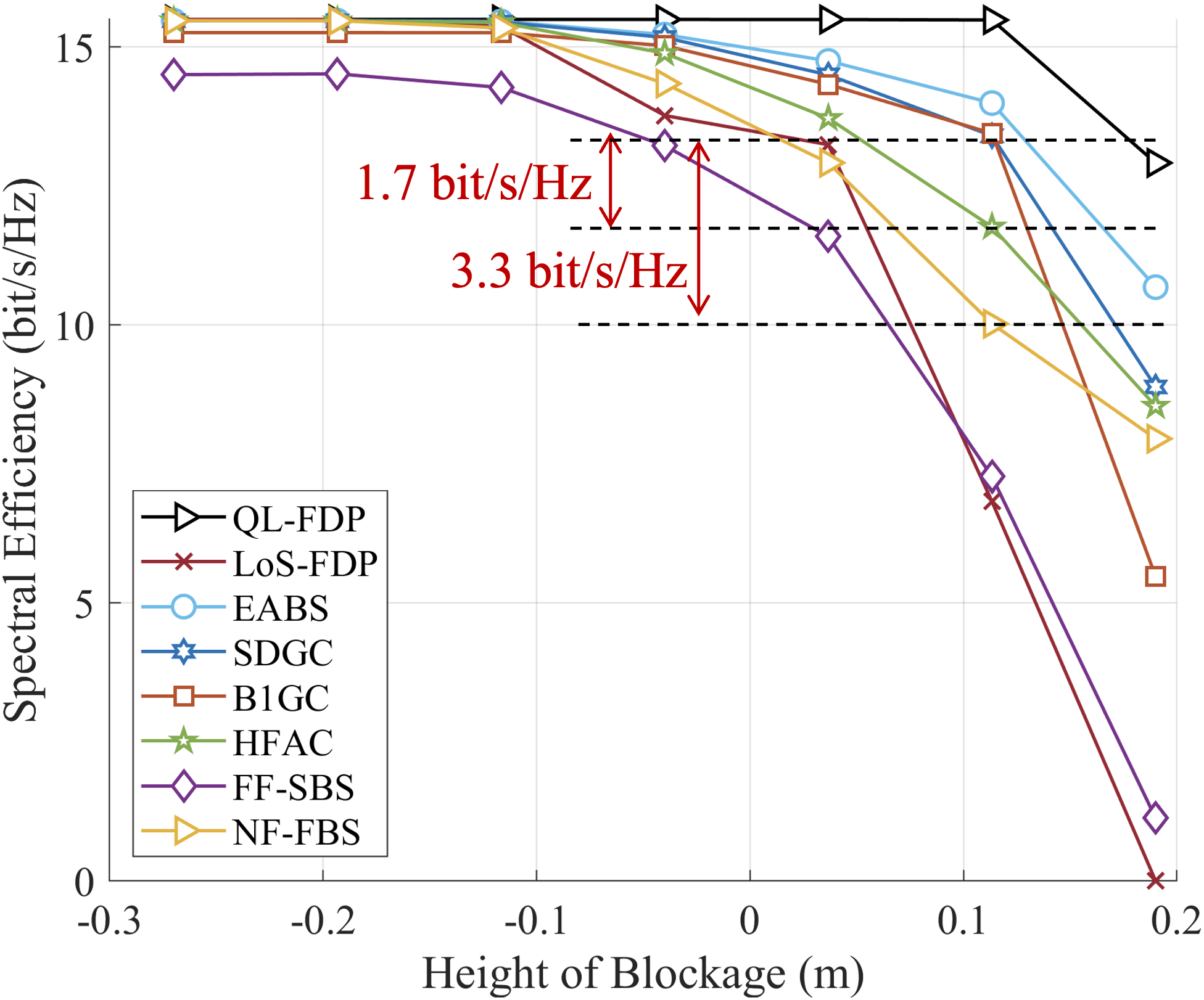} 
		\caption{Spectral efficiency versus the height of the blockage. $D=3$~m, $z_b=1.5$~m and $N_r=256$.}
		\label{fig:SE_height}
	\end{figure}
	
	\begin{figure}[t]
		\centering
		\includegraphics[width=0.8\linewidth]{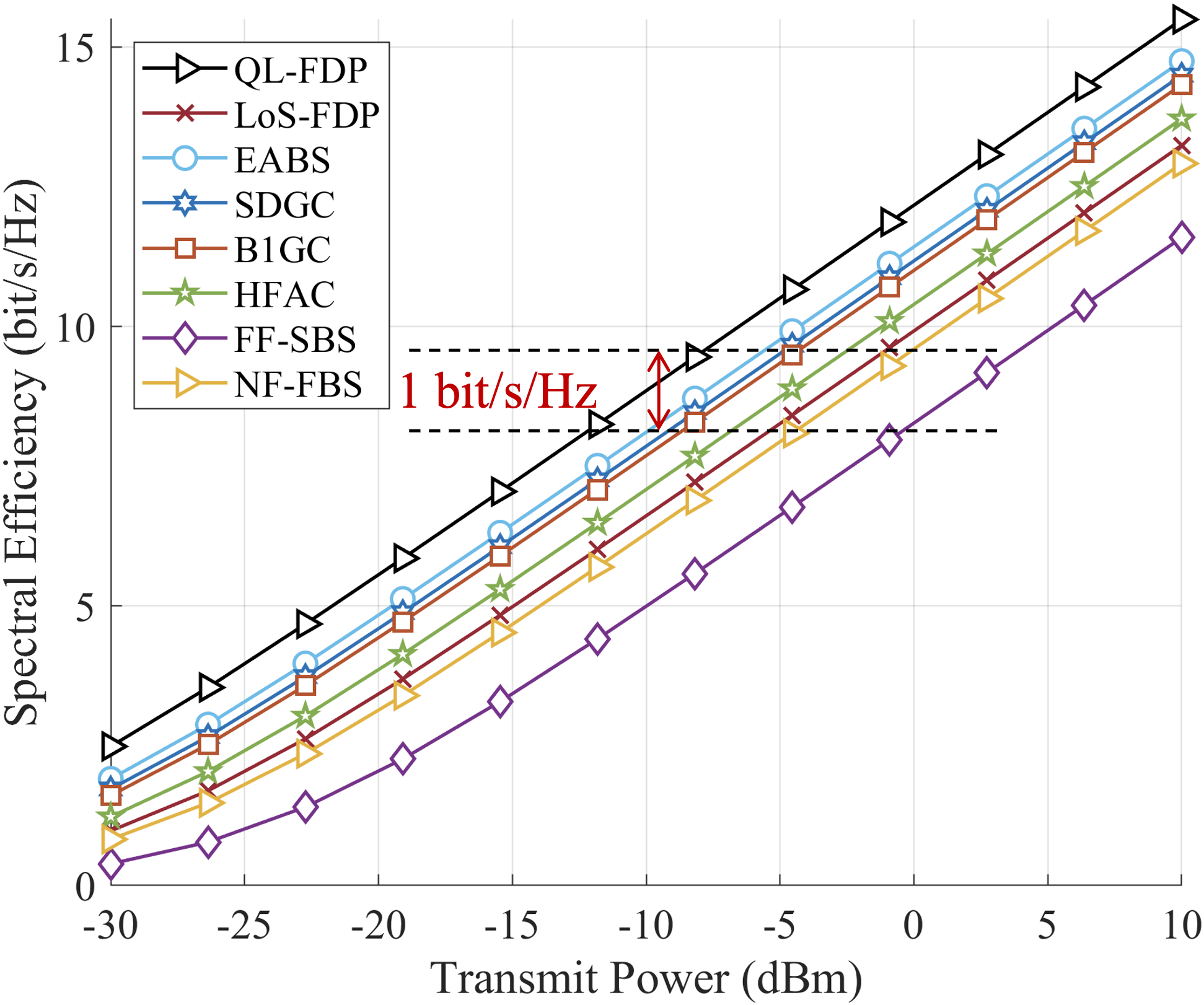} 
		\caption{Spectral efficiency versus the transmit power. $D=3$~m, $z_b=1.5$~m, $x_b=0.0367$~m and $N_r=256$.}
		\label{fig:SE_dBm}
	\end{figure}
	Fig.~\ref{fig:SE_height} evaluates the SE under different blockage heights, where the blockage is located at $z_b=1.5$~m. As the blockage becomes more severe, the LoS-FDP, FF-SBS, and NF-FBS experience substantial performance degradation, whereas the proposed SDGC and B1GC maintain significantly higher SE.
	For example, at a blockage height of approximately $-0.05$~m,
	SDGC provides approximately $1.7$~bit/s/Hz and $3.3$~bit/s/Hz
	SE gains over HFAC and NF-FBS, respectively.
	B1GC also maintains competitive SE with substantially lower
	beam training overhead. Meanwhile, SDGC remains close to the EABS benchmark over the considered blockage range, demonstrating that the proposed geometry-based codebook can achieve near-exhaustive-search performance with significantly reduced training overhead.
	Under extremely severe blockage, the performance of B1GC decreases more noticeably due to its reduced 1D search space, reflecting its intended tradeoff between spectral efficiency and training latency.

	Fig.~\ref{fig:SE_dBm} illustrates the SE versus the transmit power under the quasi-LoS scenario. As expected, the SE of all schemes increases with the transmit power. The proposed SDGC consistently achieves near-optimal performance and closely follows the EABS benchmark across the entire power range.
	At a representative transmit power of $-5$~dBm, the SE gap between SDGC and EABS is only approximately $0.13$~bit/s/Hz, while SDGC outperforms HFAC by approximately $1$~bit/s/Hz. Meanwhile, B1GC remains within approximately $0.3$~bit/s/Hz of SDGC despite its substantially lower beam training overhead. These results demonstrate that SDGC provides near-exhaustive-search performance, whereas B1GC achieves a favorable performance--latency tradeoff through its low-dimensional beam training strategy.
	In comparison, FF-SBS and NF-FBS exhibit larger performance losses because their conventional steering or focusing beam patterns lack the trajectory flexibility required for effective blockage avoidance.

	\begin{figure}[t]
		\centering
		\includegraphics[width=0.8\linewidth]{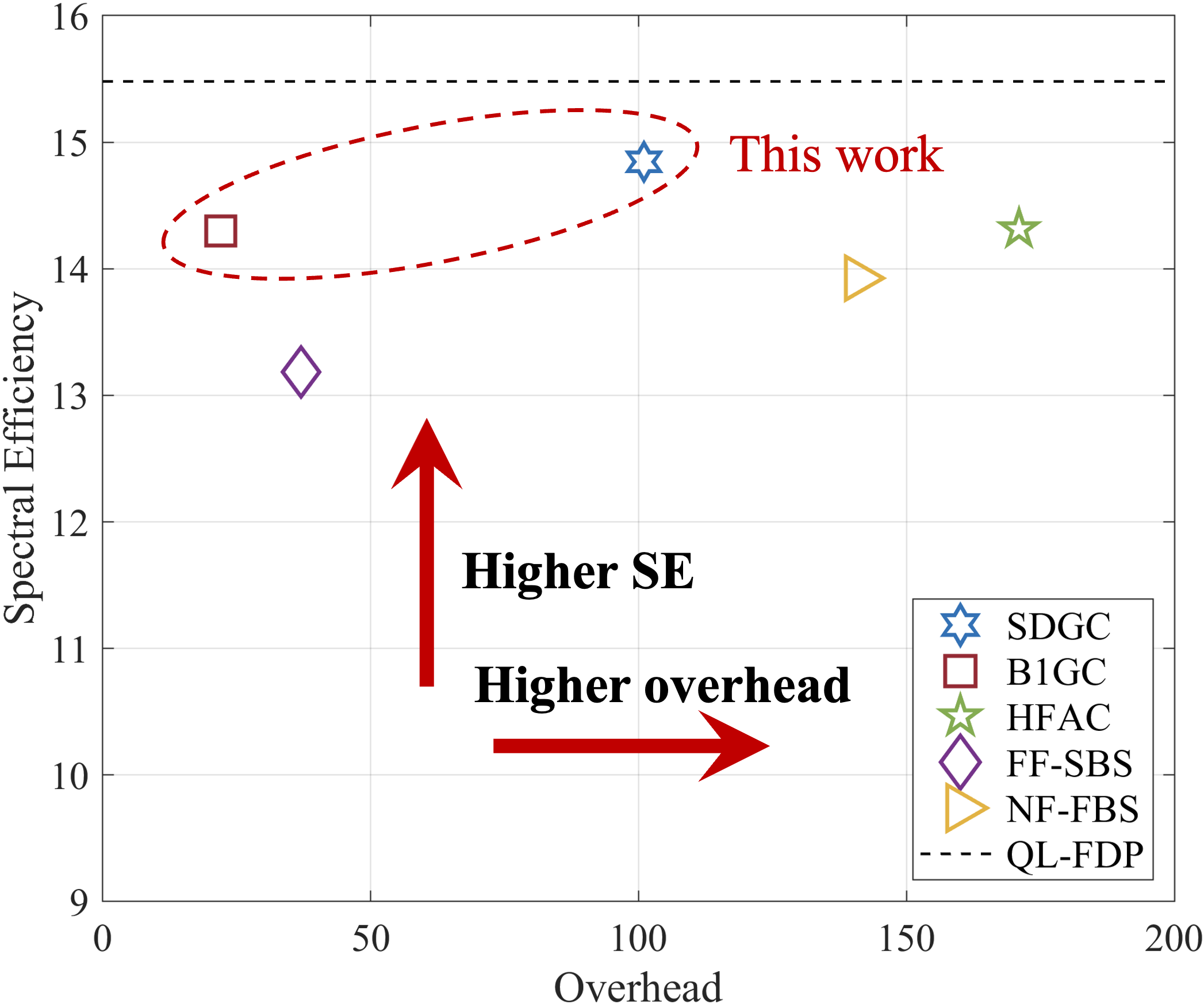} 
		\caption{Spectral efficiency versus overhead. $D=3$~m and $N_r=256$.}
		\label{fig:SE_overhead}
	\end{figure}
	To evaluate the overall robustness of different beam training strategies, Fig.~\ref{fig:SE_overhead} presents the average SE over 200 random blockage scenarios within the LoS region. The beam training overhead is measured by the total number of pilot beam transmissions, where each codeword requires one pilot and the
	two initial probe beams are included for SDGC and B1GC. Compared with HFAC~\cite{Zhao-2025-WDC}, the proposed SDGC achieves a higher average SE of approximately $14.8$~bit/s/Hz while simultaneously reducing the beam training overhead by approximately $40.9\%$. This result demonstrates the effectiveness of the proposed slope-depth geometric sampling strategy in identifying suitable Airy trajectories with reduced training overhead.
	B1GC further highlights the advantage of the proposed framework in ultra-low-latency scenarios. Specifically, B1GC reduces the beam training overhead by approximately $87.1\%$ compared with HFAC while achieving almost the same average SE. This verifies that the proposed low-dimensional geometric search can maintain robust communication performance with substantially reduced training overhead.
	For comparison, both FF-SBS and NF-FBS suffer noticeable SE degradation despite their relatively large training overhead. In particular, FF-SBS achieves approximately $13.2$~bit/s/Hz with an overhead of 37 pilots. Although NF-FBS improves the SE to approximately $13.9$~bit/s/Hz by exploiting near-field beam focusing, its performance remains inferior to the proposed Airy beam codebooks due to the lack of curved propagation capability.
	
	\begin{figure}[t]
		\centering
		\includegraphics[width=0.8\linewidth]{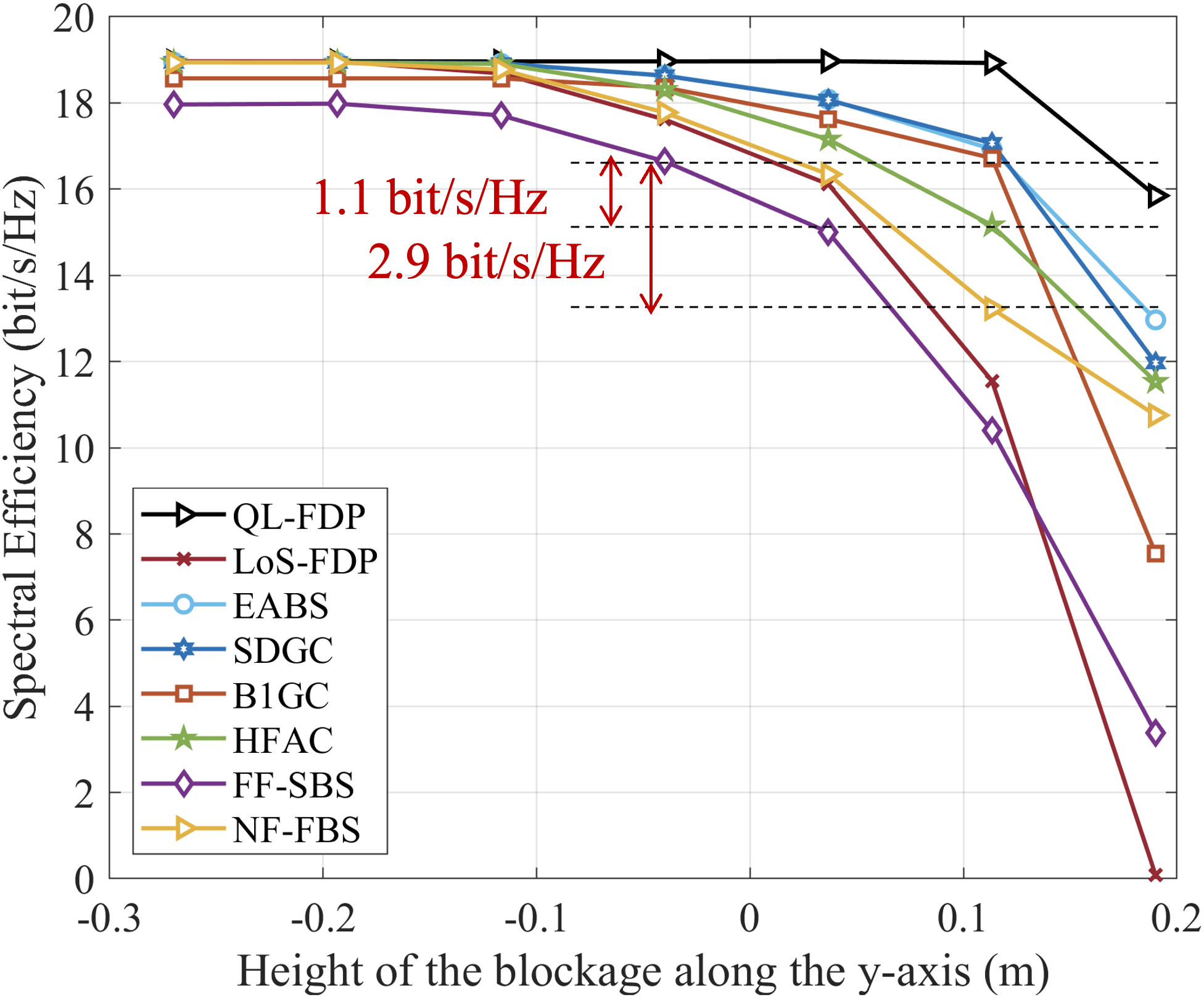}
		\caption{Spectral efficiency versus blockage upper edge in the UPA system.
			$D=3$~m, $z_b=1.5$~m, with Tx and Rx aperture sizes of
			$0.548\times0.548$~m$^2$ and $0.273\times0.273$~m$^2$, respectively.}
		\label{fig:UPA_SE}
	\end{figure}
	Finally, Fig.~\ref{fig:UPA_SE} evaluates the proposed beam training framework in the UPA system. As the blockage becomes more severe, the spectral efficiencies of LoS-FDP, FF-SBS, and NF-FBS decrease considerably due to their limited capability to circumvent the obstructed propagation region. In contrast, the proposed SDGC and B1GC maintain robust spectral efficiency and remain close to EABS over the considered blockage range. In particular, at a blockage upper edge of approximately $0.12$~m,
	SDGC provides SE gains of approximately $1.1$~bit/s/Hz and
	$2.9$~bit/s/Hz over HFAC and NF-FBS, respectively. These results demonstrate that the proposed geometry-based Airy beam training framework can be effectively extended from ULA to UPA systems through the hybrid focusing--Airy structure.

	\section{Conclusion}
	\label{sec:conclusion}
	
	In this paper, we propose a geometry-based Airy beam training framework for quasi-LoS THz near-field communications. By characterizing the theoretical feasibility boundary of Airy beam generation under finite apertures, physically infeasible beam candidates can be effectively eliminated before beam training. Based on this boundary, we develop two complementary codebook designs. The performance-oriented SDGC exploits slope-depth geometric sampling to efficiently explore the feasible waypoint region, while the latency-oriented B1GC establishes a bijective mapping between the blockage-avoidance waypoint and the receiver target to reduce the original 2D search to a 1D geometric search. The proposed framework is further extended to UPA systems through a hybrid focusing--Airy mode.
	
	Simulation results demonstrate that SDGC achieves an average SE of approximately $14.8$~bit/s/Hz while reducing the beam training overhead by $40.9\%$ compared with HFAC. B1GC further reduces the training overhead by $87.1\%$ while achieving almost the same average SE as HFAC. Compared with FF-SBS and NF-FBS, SDGC achieves approximately $1.6$ and $0.9$~bit/s/Hz higher average SE, respectively, while B1GC still provides approximately $1$ and $0.4$~bit/s/Hz gains with substantially lower training overhead. The UPA results further demonstrate that both SDGC and B1GC maintain robust performance under two-dimensional transverse blockage, validating the extension of the proposed framework from ULA to UPA systems.
	Overall, the proposed framework establishes a direct connection between geometry-based Airy wavefront design and practical beam training under unknown blockage. SDGC provides a performance-oriented solution with high spectral efficiency, whereas B1GC offers a latency-oriented alternative for rapid link establishment, enabling flexible performance--latency tradeoffs in quasi-LoS THz near-field communications.
	
	\appendices
	\section{Derivation of the Critical Feasibility Boundary}
	\label{appendix_boundary}
	
	Substituting \eqref{eq:MX_appendix_ref} into \eqref{eq:Xint_scalar_revised_short}, the tangent intercept at the receiver plane is
	\begin{equation}
		X_{\mathrm{int}}(z_r)
		=
		X+\frac{1}{16\lambda\pi^2M(X)}.
	\end{equation}
	For $B>0$, the upper feasibility boundary is reached when the tangent intercept coincides with the array edge, i.e., $X_{\mathrm{int}}(z_r)=\omega_0=\frac{D_t}{2}$,
	which yields
	\begin{equation}
		\frac{1}{16\lambda\pi^2M(X)}
		=
		\omega_0-X.
		\label{eq:appendix_rearrange}
	\end{equation}
	Substituting \eqref{eq:MX_appendix_ref} into \eqref{eq:appendix_rearrange} and simplifying, we obtain
	\begin{equation}
		1
		=
		(\omega_0-X)
		\left[
		\frac{3X}{\omega_0^2}
		+
		\sqrt{
			\frac{9X^2}{\omega_0^4}
			+
			\frac{6}{\omega_0^2}
		}
		\right].
		\label{eq:appendix_simplified}
	\end{equation}
	Rearranging \eqref{eq:appendix_simplified} gives
	\begin{equation}
		\frac{1}{\omega_0-X}
		-
		\frac{3X}{\omega_0^2}
		=
		\sqrt{
			\frac{9X^2}{\omega_0^4}
			+
			\frac{6}{\omega_0^2}
		}.
	\end{equation}
	Squaring both sides and simplifying lead to
	\begin{equation}
		\omega_0^2
		-
		6X(\omega_0-X)
		=
		6(\omega_0-X)^2.
	\end{equation}
	After straightforward algebraic manipulation, the quadratic terms cancel and we obtain
	\begin{equation}
		X^\star
		=
		\frac{5}{6}\omega_0
		=
		\frac{5}{12}D_t.
	\end{equation}

	\section{Proof of Lemma 1}
	\label{app:lemma_proof}
	
	We begin with the analytical trajectory of the main lobe in \eqref{eq:traj}, given by
	\begin{equation}
		x(z)
		=
		-\xi_{\mathrm{peak}}\lambda Bz-\sin\theta z
		-\frac{S_R^2-S_I^2}{16\lambda\pi^2B^3}z,
	\end{equation}
	where $S_R=1/z-1/F$ and $S_I=\lambda/(\pi\omega_0^2)$. Since
	\begin{align}
		(S_R^2-S_I^2)z
		&=
		z\left[
		\frac{1}{z^2}
		-\frac{2}{Fz}
		+\frac{1}{F^2}
		-S_I^2
		\right] \nonumber\\
		&=
		\frac{1}{z}
		-\frac{2}{F}
		+
		\left(
		\frac{1}{F^2}-S_I^2
		\right)z,
	\end{align}
	the trajectory can be reorganized as
	\begin{equation}
		x(z)=C_1z+C_0-\frac{\alpha}{z},
		\label{eq:app_traj_standard}
	\end{equation}
	where
	\begin{align}
		C_1
		&=
		-\xi_{\mathrm{peak}}\lambda B-\sin\theta
		-\frac{1}{16\lambda\pi^2B^3}
		\left[
		\frac{1}{F^2}
		-
		\left(
		\frac{\lambda}{\pi\omega_0^2}
		\right)^2
		\right],
	\end{align}
	$C_0=1/(8\lambda\pi^2B^3F)$, and
	$\alpha=1/(16\lambda\pi^2B^3)$.
	
	Next, define the secant line connecting the near-field alert point $(z_p,x_p)$ and the receiver point $(z_r,x_r)$ as
	\begin{equation}
		L(z)=x_p+\frac{x_r-x_p}{z_r-z_p}(z-z_p).
	\end{equation}
	Using the boundary conditions $x_p=C_1z_p+C_0-\alpha/z_p$ and
	$x_r=C_1z_r+C_0-\alpha/z_r$, the difference between the Airy trajectory and the LoS secant can be simplified as
	\begin{align}
		x(z)-L(z)
		&=
		\left(
		C_1z+C_0-\frac{\alpha}{z}
		\right)
		-
		\Bigg[
		C_1z_p+C_0-\frac{\alpha}{z_p}
		\nonumber\\
		&\quad+
		\frac{
			\left(C_1z_r-\alpha/z_r\right)
			-
			\left(C_1z_p-\alpha/z_p\right)}
		{z_r-z_p}
		(z-z_p)
		\Bigg]
		\nonumber\\
		&=
		\alpha
		\left[
		\frac{z-z_p}{zz_p}
		+
		\frac{1/z_r-1/z_p}{z_r-z_p}(z-z_p)
		\right]
		\nonumber\\
		&=
		\alpha
		\frac{(z-z_p)(z_r-z)}{zz_pz_r}.
		\label{eq:app_bulge}
	\end{align}
	For $z\in[z_p,z_r]$, the term $(z-z_p)(z_r-z)$ is positive, which indicates that the Airy trajectory bends outward relative to the LoS secant path. Therefore, the enveloped spatial area is
	\begin{align}
		S
		&=
		\int_{z_p}^{z_r}
		\alpha
		\frac{
			-z^2+(z_p+z_r)z-z_pz_r}
		{zz_pz_r}
		dz \nonumber\\
		&=
		\alpha
		\left[
		\frac{z_r^2-z_p^2}{2z_pz_r}
		-
		\ln\left(\frac{z_r}{z_p}\right)
		\right].
		\label{eq:app_area}
	\end{align}
	
	According to the optimal parameter derivation in \eqref{eq:closed_form_B}, the bending parameter satisfies $B^3\propto\mathcal{O}(1/z_p)$ as $z_p\rightarrow0$. Since $\alpha=1/(16\lambda\pi^2B^3)$, there exists a positive constant $K$ such that $\alpha=Kz_p$. Substituting this relation into \eqref{eq:app_area} gives
	\begin{equation}
		S(z_p)
		=
		K\frac{z_r}{2}
		-
		K\frac{z_p^2}{2z_r}
		-
		Kz_p\ln\left(\frac{z_r}{z_p}\right).
		\label{eq:app_area_asymptotic}
	\end{equation}
	Finally, the logarithmic penalty term satisfies
	\begin{align}
		\lim_{z_p\to0^+}
		z_p\ln\left(\frac{z_r}{z_p}\right)
		&=
		\lim_{z_p\to0^+}
		\frac{\ln(z_r/z_p)}{1/z_p}
		=
		\lim_{z_p\to0^+}z_p
		=0,
	\end{align}
	where L'Hôpital's rule is applied. Hence,
	\begin{equation}
		\lim_{z_p\to0^+}S(z_p)=K\frac{z_r}{2},
	\end{equation}
	which corresponds to the maximum achievable spatial coverage. Moreover, for $z_p<z_r/e$, we have $S'(z_p)<0$, indicating that the enveloped spatial area monotonically increases as the reference depth $z_p$ decreases. This completes the proof.

	\section{Proof of Lemma~\ref{lemma:coverage}}
	\label{app:coverage_proof}
	
	Let the transverse position of the Airy-beam main lobe at propagation distance $z\in(z_f,z_r)$ be denoted by
	\begin{equation}
		x(z)=\mathcal{H}(z;B,F,\theta).
	\end{equation}
	Since the beam parameters $\{B,F,\theta\}$ are closed-form functions of $(x_s,x_r)$, and $x_r$ is linearly determined by $x_s$ through \eqref{eq:1D_mapping}, the trajectory can be equivalently expressed as $x(z;x_s)$.
	
	Because the trajectory equation and the closed-form beam parameters consist of continuous algebraic and trigonometric operations within the feasible domain, the mapping $x_s \mapsto x(z;x_s)$
	is continuous for any fixed $z\in(z_f,z_r)$.
	Therefore, for any transverse position $x_0$ lying between
	$x(z;x_{s,\min})$ and $x(z;x_{s,\max})$, the Intermediate Value Theorem guarantees the existence of at least one
	$x_s^\star\in[x_{s,\min},x_{s,\max}]$ such that
	$x(z;x_s^\star)=x_0$.
	Hence, at every propagation distance $z\in(z_f,z_r)$, the continuous family of Airy trajectories covers all transverse positions between the two boundary trajectories corresponding to $x_{s,\min}$ and $x_{s,\max}$. Consequently, the generated trajectory family continuously covers the spatial region bounded by these two trajectories without introducing spatial blind spots.

	\newpage


\begin{thebibliography}{10}
\providecommand{\url}[1]{#1}
\csname url@samestyle\endcsname
\providecommand{\newblock}{\relax}
\providecommand{\bibinfo}[2]{#2}
\providecommand{\BIBentrySTDinterwordspacing}{\spaceskip=0pt\relax}
\providecommand{\BIBentryALTinterwordstretchfactor}{4}
\providecommand{\BIBentryALTinterwordspacing}{\spaceskip=\fontdimen2\font plus
\BIBentryALTinterwordstretchfactor\fontdimen3\font minus \fontdimen4\font\relax}
\providecommand{\BIBforeignlanguage}[2]{{%
\expandafter\ifx\csname l@#1\endcsname\relax
\typeout{** WARNING: IEEEtran.bst: No hyphenation pattern has been}%
\typeout{** loaded for the language `#1'. Using the pattern for}%
\typeout{** the default language instead.}%
\else
\language=\csname l@#1\endcsname
\fi
#2}}
\providecommand{\BIBdecl}{\relax}
\BIBdecl

\bibitem{Akyildiz-2022-Terahertz}
I.~F. Akyildiz, C.~Han, Z.~Hu, S.~Nie, and J.~M. Jornet, ``{Terahertz Band Communication: An Old Problem Revisited and Research Directions for the Next Decade},'' \emph{IEEE Trans. Commun.}, vol.~70, no.~6, pp. 4250--4285, Jun. 2022.

\bibitem{Chen-2019-A-survey}
Z.~Chen, X.~Ma, B.~Zhang, Y.~Zhang, Z.~Niu, N.~Kuang, W.~Chen, L.~Li, and S.~Li, ``{A Survey on Terahertz Communications},'' \emph{China Commun.}, vol.~16, no.~2, pp. 1--35, Mar. 2019.

\bibitem{Han-2021-Hybrid}
C.~Han, L.~Yan, and J.~Yuan, ``{Hybrid Beamforming for Terahertz Wireless Communications: Challenges, Architectures, and Open Problems},'' \emph{IEEE Wireless Commun.}, vol.~28, no.~4, pp. 198--204, Aug. 2021.

\bibitem{An-2024-Near-Field}
J.~An, C.~Yuen, L.~Dai, M.~Di~Renzo, M.~Debbah, and L.~Hanzo, ``{Near-Field Communications: Research Advances, Potential, and Challenges},'' \emph{IEEE Wireless Commun.}, vol.~31, no.~3, pp. 100--107, Jun. 2024.

\bibitem{Yuan-2023-Spatial}
Z.~Yuan, J.~Zhang, Y.~Ji, G.~F. Pedersen, and W.~Fan, ``{Spatial Non-Stationary Near-Field Channel Modeling and Validation for Massive MIMO Systems},'' \emph{IEEE Trans. Antennas Propag.}, vol.~71, no.~1, pp. 921--933, Jan. 2023.

\bibitem{Han-2025-datacenter-mag}
C.~Han, M.~Zhu, W.~Zhao, Z.~Yu, G.~Huang, G.~Wang, Z.~Chen, W.~Tong, and W.~Zhang, ``{When Wires Can’t Keep Up: Reconfigurable AI Data Centers Empowered by Terahertz Wireless Communications},'' \emph{IEEE Wireless Commun.}, 2026.

\bibitem{Ayach-2014-Spatially}
O.~E. Ayach, S.~Rajagopal, S.~Abu-Surra, Z.~Pi, and R.~W. Heath, ``{Spatially Sparse Precoding in Millimeter Wave MIMO Systems},'' \emph{IEEE Trans. Wireless Commun.}, vol.~13, no.~3, pp. 1499--1513, Mar. 2014.

\bibitem{Alkhateeb-2015-Limited}
A.~Alkhateeb, G.~Leus, and R.~W. Heath, ``{Limited Feedback Hybrid Precoding for Multi-User Millimeter Wave Systems},'' \emph{IEEE Trans. Wireless Commun.}, vol.~14, no.~11, pp. 6481--6494, Nov. 2015.

\bibitem{Zhang-2022-Beam-focusing}
H.~Zhang, N.~Shlezinger, F.~Guidi, D.~Dardari, M.~F. Imani, and Y.~C. Eldar, ``{Beam Focusing for Near-Field Multiuser MIMO Communications},'' \emph{IEEE Trans. Wireless Commun.}, vol.~21, no.~9, pp. 7476--7490, Sep. 2022.

\bibitem{Cui-2022-Channel-estimation}
M.~Cui and L.~Dai, ``{Channel Estimation for Extremely Large-Scale MIMO: Far-Field or Near-Field?}'' \emph{IEEE Trans. Commun.}, vol.~70, no.~4, pp. 2663--2677, Apr. 2022.

\bibitem{Cui-2023-Near-Field-MIMO}
M.~Cui \emph{et~al.}, ``{Near-Field MIMO Communications for 6G: Fundamentals, Challenges, Potentials, and Future Directions},'' \emph{IEEE Commun. Mag.}, vol.~61, no.~1, pp. 40--46, Jan. 2023.

\bibitem{Wu-2023-Multiple}
Z.~Wu and L.~Dai, ``{Multiple Access for Near-Field Communications: SDMA or LDMA?}'' \emph{IEEE J. Sel. Areas Commun.}, vol.~41, no.~6, pp. 1918--1935, Jun. 2023.

\bibitem{Nikolaos-2019-Airy}
N.~K. Efremidis, Z.~Chen, M.~Segev, and D.~N. Christodoulides, ``{Airy Beams and Accelerating Waves: An Overview of Recent Advances},'' \emph{Optica}, vol.~6, no.~5, pp. 686--701, May 2019.

\bibitem{Zhan-2020-Propagations}
K.~Zhan, W.~Zhang, R.~Jiao, L.~Dou, and B.~Liu, ``{Propagations of Airy Beams with Quadratic Phase Modulation, and Their Interaction in Paraxial Optical Systems},'' \emph{Opt. Commun.}, vol. 474, p. 126156, Nov. 2020.

\bibitem{Jornet-2023-Wireless}
J.~M. Jornet, E.~Knightly, and D.~Mittleman, ``{Wireless Communications Sensing and Security Above 100 GHz},'' \emph{Nat. Commun.}, vol.~14, no. 841, 2023.

\bibitem{Guerboukha-2024-Curving}
H.~Guerboukha, B.~Zhao, Z.~Fang \emph{et~al.}, ``{Curving THz wireless data links around obstacles},'' \emph{Commun. Eng.}, vol.~3, no.~58, pp. 1--1, 2024.

\bibitem{Chen-2025-physics}
{H. Chen, A. Kludze and Y. Ghasempour}, ``{A physics-informed Airy Beam Learning Framework for Blockage Avoidance in Sub-terahertz Wireless Networks.}'' \emph{Nat. Commun.}, vol.~16, no. 7387, Aug. 2025.

\bibitem{Chen-2024-Curving}
H.~Chen, A.~Kludze, and Y.~Ghasempour, ``{Curving Around Obstacles via NN-Enabled Wavefront Shaping in Sub-THz Wireless Networks},'' in \emph{Proc. of IEEE Globecom}, Dec. 2024.

\bibitem{Liu-2025-Near-Field-THz-Bending}
A.~Liu, W.~Mei, P.~Wang, D.~Wang, Y.~F. Wu, Z.~Chen, and B.~Ning, ``{Near-Field THz Bending Beamforming: A Convex Optimization Perspective},'' in \emph{2025 IEEE 102nd Vehicular Technology Conference (VTC2025-Fall)}, 2025, pp. 1--5.

\bibitem{Lu-2026-Analytical}
Z.~Lu, Y.~Han, R.~Sun, J.~Nie, and S.~Jin, ``{Analytical Framework of Airy Beams in Near-Field XL-MIMO: From Ideal Optics to Wireless Reality},'' \emph{arXiv:2605.08655}, May 2026.

\bibitem{Wang-2026-Cubic}
Y.~Wang and L.~Dai, ``Why is cubic-phase {Airy} beamforming sufficient for blockage recovery?'' \emph{arXiv:2609.01313}, Sep. 2026.

\bibitem{Zhao-2026-Airy}
W.~Zhao, C.~Han, and E.~Bj{\"o}rnson, ``{Airy Beam Engineering in Near-field Communications: A Tractable Closed-Form Analysis in the Terahertz Band},'' \emph{arXiv:2603.13866}, Mar. 2026.

\bibitem{Wang-2026-Physics-Guided}
Y.~Wang and L.~Dai, ``Physics-guided neural {Airy} beamforming for near-field blockage mitigation,'' \emph{arXiv:2608.04388}, Aug. 2026.

\bibitem{Petrov-2024-Wavefront}
V.~Petrov, H.~Guerboukha, A.~Singh, and J.~M. Jornet, ``{Wavefront Hopping for Physical Layer Security in 6G and Beyond Near-Field THz Communications},'' \emph{IEEE Trans. Commun.}, vol.~73, no.~5, pp. 2996--3012, May 2025.

\bibitem{Petrov-2024-Wavefront-Hopping}
V.~Petrov, H.~Guerboukha, D.~M. Mittleman, and A.~Singh, ``{Wavefront Hopping: An Enabler for Reliable and Secure Near Field Terahertz Communications in 6G and Beyond},'' \emph{IEEE Wireless Commun.}, vol.~31, no.~1, pp. 48--55, Feb. 2024.

\bibitem{Singh-2024-Wavefront}
A.~Singh, V.~Petrov, H.~Guerboukha, I.~V. Reddy, E.~W. Knightly, D.~M. Mittleman, and J.~M. Jornet, ``{Wavefront Engineering: Realizing Efficient Terahertz Band Communications in 6G and Beyond},'' \emph{IEEE Wireless Commun.}, vol.~31, no.~3, pp. 133--139, Jan. 2024.

\bibitem{Weng-2025-Learning}
C.~Weng, Y.~Guo, B.~Zhao, Y.~Wang, W.~Chen, and Z.~Li, ``{Learning-Based Blockage-Resilient Beam Training in Near-Field Terahertz Communications},'' \emph{IEEE Trans. Wireless Commun.}, vol.~25, pp. 20\,945--20\,961, 2026.

\bibitem{Zhao-2025-WDC}
W.~Zhao, S.~Abadal, G.~Song, J.~Jiang, and C.~Han, ``{Terahertz Wireless Data Center: Gaussian Beam or Airy Beam?}'' \emph{IEEE Trans. Wireless Commun.}, vol.~25, pp. 7922--7938, 2026.

\bibitem{zhang-2026-breaking}
S.~Zhang, B.~Di, and L.~Song, ``{Breaking Near-Field Communication Barriers: Focused, Curved, or Airy Beamforming?}'' \emph{arXiv:2604.01704}, Apr. 2026.

\bibitem{zhang-2026-hybrid}
------, ``{Hybrid Digital and Analog Airy Beamforming for Near-Field Multi-User Communications},'' \emph{arXiv:2605.29481}, May 2026.

\bibitem{Han-2015-Multi-Ray}
C.~Han, A.~O. Bicen, and I.~F. Akyildiz, ``{Multi-Ray Channel Modeling and Wideband Characterization for Wireless Communications in the Terahertz Band},'' \emph{IEEE Trans. Wireless Commun.}, vol.~14, no.~5, pp. 2402--2412, May 2015.

\end{thebibliography}
\end{document}